%% file: main.tex
\documentclass{egpubl}
\ifdefined\pdfminorversion\pdfminorversion=7\fi

\preprint
\copyrightTextTitPag{}
\copyrightTextRunPag{}
\pdfSubject{Composable CAD: Sketch-to-CAD Program Synthesis in Parts}
\makeatletter
\def\author@nextline{%
  \tabularnewline[11pt]\reset@font\small\rmfamily
  \let\\=\author@nextlinetwo\ignorespaces
}
\def\ps@titlepage{%
  \let\@mkboth\@gobbletwo
  \def\@oddhead{}%
  \let\@evenhead\@oddhead
  \def\@oddfoot{\hfil\thepage\hfil}%
  \let\@evenfoot\@oddfoot
  \let\sectionmark\EmptySectionmark
  \let\subsectionmark\EmptySubsectionmark
}
\makeatother

\usepackage[T1]{fontenc}
\usepackage{dfadobe}
\usepackage{amsmath,amssymb}
\usepackage{tabularx,booktabs}
\newtheorem{proposition}{Proposition}
\usepackage{cite}
\BibtexOrBiblatex
\electronicVersion
\PrintedOrElectronic
\usepackage{graphicx}
\usepackage{egweblnk}
\usepackage{soul,placeins,xcolor,flushend}
\usepackage[ruled,vlined,linesnumbered]{algorithm2e}
\usepackage{rotating}
\hypersetup{
  pdfsubject={Sketch-to-CAD program synthesis},
  pdfkeywords={CAD, sketch-based modeling, program synthesis, composable design}
}

\title[Composable CAD]{Composable CAD: Sketch-to-CAD Program Synthesis in Parts}
\author[Y. Li et al.]
{%
  \parbox{\textwidth}{\centering
    Yuanbo Li,
    Vivian Liu$^{*}$,
    Sean J. Liu$^{*}$,
    and Justin Matejka
  }%
  \\
  \parbox{\textwidth}{\centering
    Autodesk Research\\
    San Francisco, CA, US\\
    $^{*}$Equal contribution
  }%
}

\begin{document}

\teaser{
\includegraphics[width=1.0\linewidth]{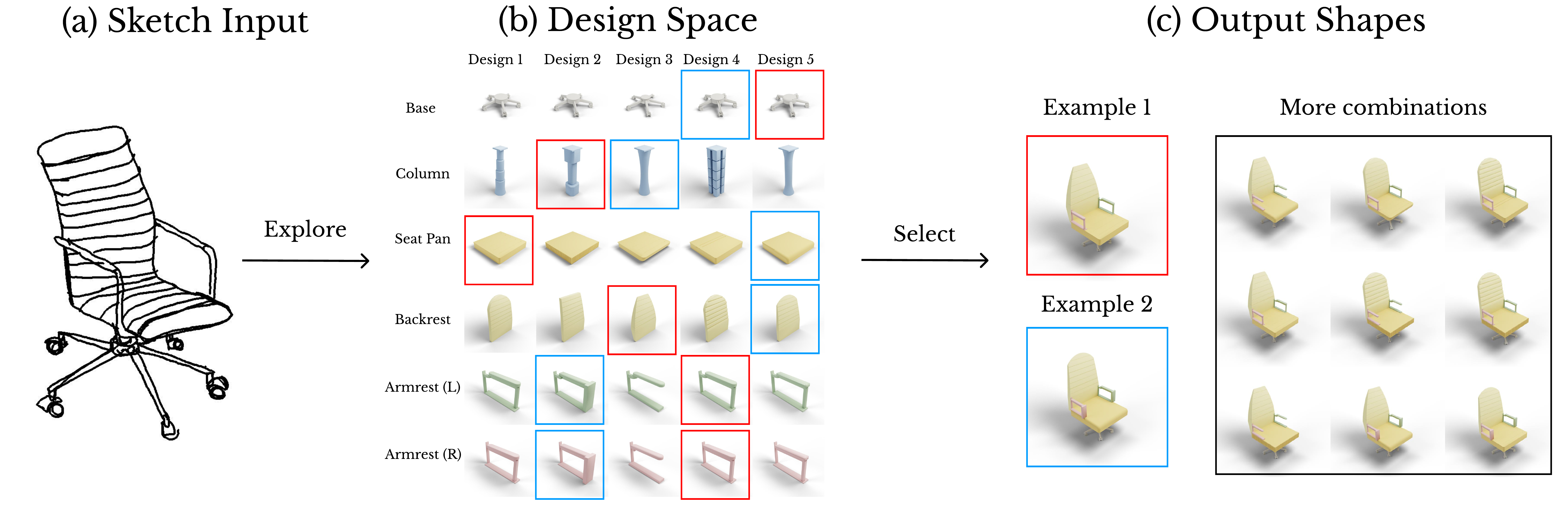}
 \centering
\caption{Given one freehand sketch (a), our system segments the object into parts and generates several samples for each part (b). The user composes a shape by selecting any sample per part (Example 1 in red, Example 2 in blue), and different selections compose into different shapes (c). The output is a CadQuery program for each part.}
\label{fig:teaser}
}

\maketitle
\hypersetup{
  pdftitle={Composable CAD: Sketch-to-CAD Program Synthesis in Parts},
  pdfauthor={Yuanbo Li, Vivian Liu, Sean J. Liu, and Justin Matejka}
}

\begin{abstract}
Designers often build CAD models part by part, and may iterate on individual components throughout the design process. Recent vision-language models (VLMs) can synthesize CAD programs from sketches, but generating parts independently does not ensure that they are composable: components may lose the intended proportions and alignments or fail to join correctly where they meet. We present Composable CAD, a framework that uses off-the-shelf VLMs to synthesize a design space of independently sampled, interchangeable component programs, where any selection composes into a coherent CAD model. We recover design constraints from the sketch and use them to construct a coarse proxy that establishes how the components are sized, aligned, and attached. We then generate each component independently from its proxy and apply a deterministic rewrite to ensure that neighboring components meet at their prescribed interfaces. We evaluate Composable CAD across a range of objects and sketch styles and conduct two user studies. Compared to VLM-only baselines, our method produces more detailed, composable designs.

\end{abstract}

\input{sections/1_Intro}
\input{sections/2_RelatedWorks}
\input{sections/3_Overview}

\input{sections/4_DeterministicAlgorithm}
\input{sections/6_Applications}
\input{sections/7_Evaluations}
\input{sections/8_Conclusion}

\clearpage

\input{sections/fig_only}

\clearpage

\bibliographystyle{eg-alpha-doi}
\bibliography{references}

\clearpage

\appendix
\input{sections/9_Appendix}

\end{document}

%% file: sections/1_Intro.tex
\section{Introduction}
\label{sec:intro}

 In industrial design workflows, designers typically begin with a sketch~\cite{kwon2024sketch2prototype, ban2019sketchsynthesis}. The sketch records the initial concept: which parts make up the design, how those parts are laid out in space, and what details set the design apart. Sketching serves as the front end of ideation, letting designers visualize and test ideas rapidly~\cite{eissen2007sketching, buxton2007sketching}. Once designers settle on a concept, they rebuild it in a CAD tool at higher fidelity and explore more precise variations of it~\cite{schulz2017interactive, li2024comparative}.

Recent vision-language models (VLMs) attempt to automate CAD generation from sketches~\cite{anthropic2026opus48,geminiteam2025gemini25}. A designer can provide a foundation model an input text prompt ``generate CAD for a chair'' and sketch and receive a CAD program as output ~\cite{anthropic2026opus48, geminiteam2025gemini25}. A growing body of work in this space instantiates agentic coding loops where the model executes its CAD programs, inspects renders for visual feedback, and revises the CAD geometry over several rounds~\cite{Alrashedy2024CADCodeVerify, Mallis2025CADAssistant, Ocker2025IdeaToCAD, Fan2025CADDesigner, Barkley2026CADSmith, ZeroToCAD2026}.

\begin{figure}[t!]
  \includegraphics[width=\linewidth]{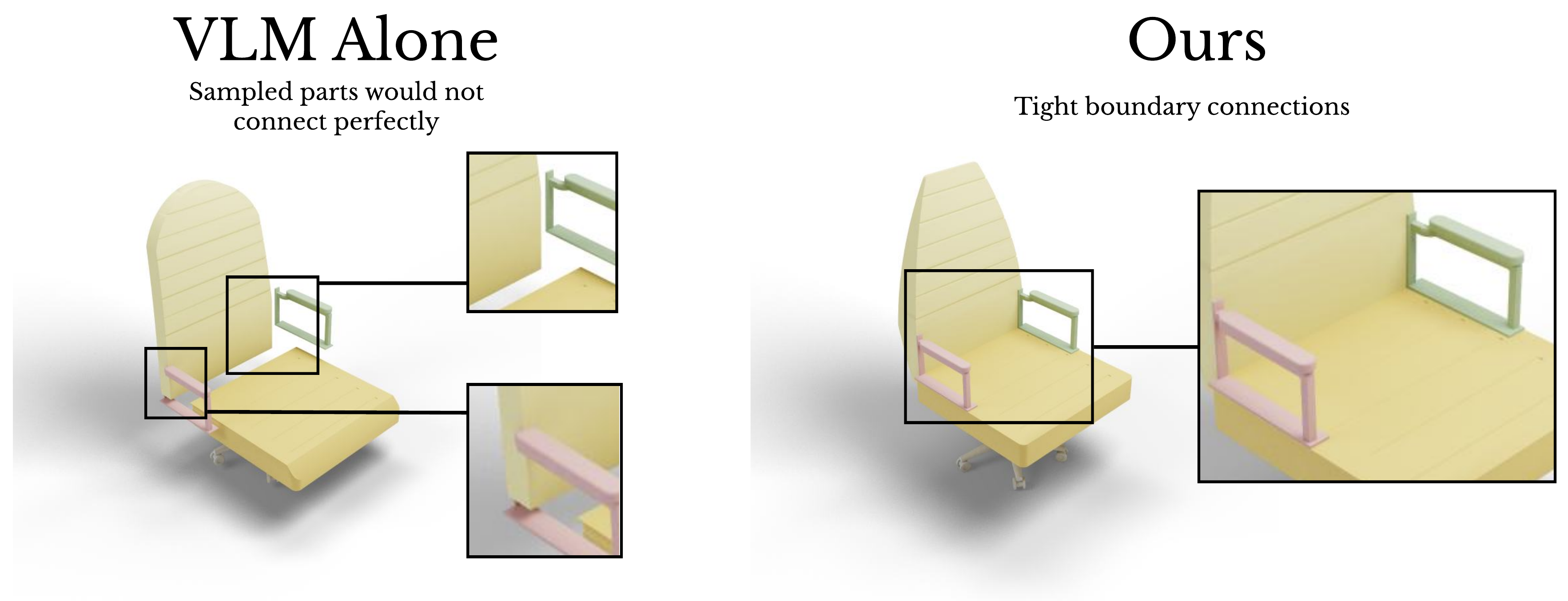}
\caption{Independently sampled parts do not connect on their own. When each part is generated in a separate VLM call, with no communication between the calls, the parts leave gaps and misalignments where they should meet (left; insets show the junctions between armrest and seat and between armrest and backrest failing to close). Our algorithm closes these interfaces, so adjacent parts meet congruently with no gap or overlap (right).}   \label{fig:compare_sketchSoup}
\end{figure}

However, independently generated parts which may be valid in isolation, are not necessarily \textit{composable}: the parts may not fit together, fail to join without gaps or overlaps, and have hallucinated away from the original proportions, alignments, and design intents defined in the original sketch (Figure~\ref{fig:compare_sketchSoup}). Guaranteeing composability in CAD generation raises a few challenges. First, the agent must preserve the designer-set constraints -- the part decomposition, proportions, and alignments laid out in the sketch. Second, the agent must preserve the structural constraints and relationships between parts, so they can join without gaps or volumetric overlap. This is difficult for VLMs because the surfaces where parts meet are often implicit in CAD programs and emerge only when the operations execute. An extrusion, for example, creates surfaces with no explicit handle in the code. Additionally, structured exploration can give users choices for how to refine the generated CAD assembly.

We present \emph{Composable CAD}, a framework that turns a concept sketch into a design space of independently sampled, composable part programs (Figure~\ref{fig:teaser}). Designers explore alternatives for each part and compose a design from their selections. Sketch-derived relations guide part proportions, alignments, and attachments, while a deterministic rewriting algorithm enforces structural compatibility across the sampled variants.

Our system follows a two-stage, coarse-to-fine pipeline. In the planning stage, we recover attachment, alignment, and proportion relations from the sketch and use them to construct a coarse proxy that fixes the layout of the parts. In the detail generation stage, we elaborate each part independently from its proxy. Finally, we use a deterministic rewrite algorithm to reconcile independently generated parts at their connecting surfaces, making them interchangeable across samples.

We demonstrate Composable CAD across a variety of objects and sketch styles, producing composable, part-based design spaces. Compared with VLM baselines that generate objects as a whole or create parts separately, our method produces detailed parts while maintaining composability across variants. We conducted two user studies: one comparing the perceptual quality of our outputs to a single-pass VLM baseline, and another comparing our part-based exploration workflow to a VLM-only baseline. Our results show that participants preferred our generated shapes and favored our part-based workflow of selecting and combining individual parts.

In summary, our contributions are:
\begin{itemize}
\item A formulation of sketch-to-CAD synthesis as a design space of independently sampled, interchangeable part programs, and its application to exploratory design.
\item A coarse-to-fine generation pipeline that recovers geometric relations from the sketch, plans the object as a shared proxy, and generates each part independently.
\item A deterministic rewrite algorithm that localizes connecting surfaces and edits independently generated part programs to meet congruently.
\end{itemize}

%% file: sections/2_RelatedWorks.tex
\section{Related Works}
\label{sec:related_works}

\begin{figure*}[t!]
\includegraphics[width=\linewidth, height=5cm]{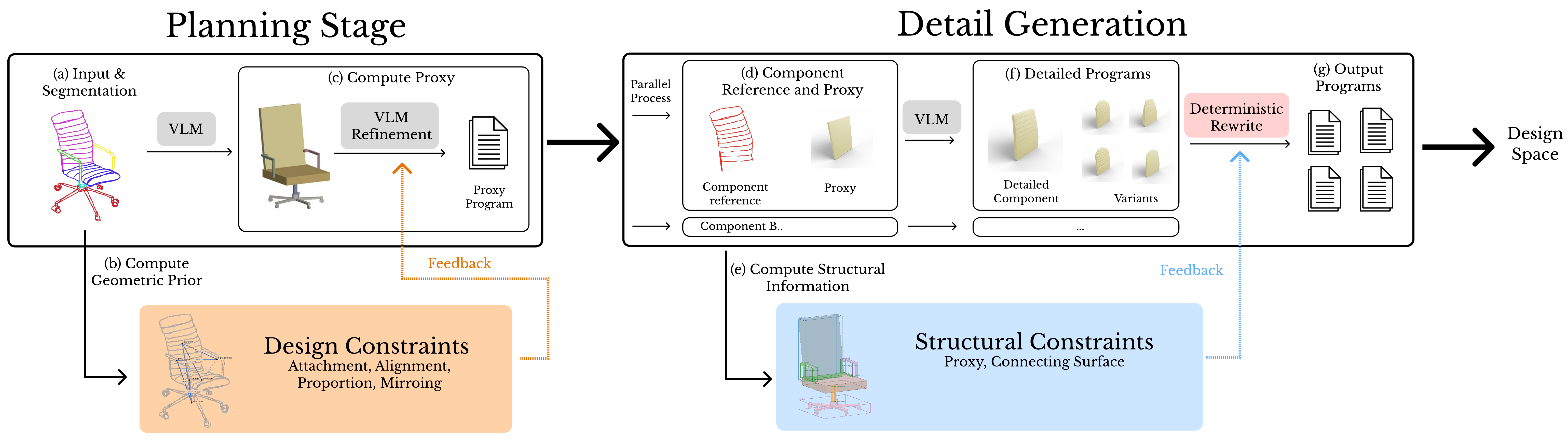}
\caption{System pipeline. Planning segments the sketch (a), extracts design constraints from stroke geometry (b), and refines a proxy program (c), which defines the structural constraints (e). Detail generation uses each part's proxy program and corresponding sketch region (d) to produce detailed variants in parallel (f). A deterministic rewrite enforces the structural constraints (g), making any selection of part variants compatible.}\label{fig:pipeline}
\end{figure*}

Our work builds on three areas of research: sketch-based modeling, CAD program synthesis, and part-based shape generation. We refer to recent surveys for broader discussions of these domains~\cite{LiuSurvey, Neurosymbolic2023}.

\paragraph*{Sketch-based modeling.}

Generating 3D geometry from 2D sketches has a long history. Early methods reconstruct 3D shapes from line drawings using geometric regularities such as parallelism, orthogonality, planarity, and symmetry~\cite{LIPSON1996Optimisation,liu2008plane,yang2013complex,Cordier2013,PLUMED20161}. Later methods handle curved geometry, over-sketching, and construction lines~\cite{True2Form,schmidt2009analytic,ILoveSketch,sadri2014flow,Uplift3D,Symmetry3D,Yu:2021:ScaffoldSketch}. We similarly exploit geometric cues in sketch strokes, but use them to constrain VLM-generated CAD rather than reconstruct geometry directly.

We contribute to the family of works that infer parametric shapes or CAD operations from sketches~\cite{huang2016shape,NGGBB16,Sketch2CAD,Free2CAD,Puhachov2023,sketch2seq,li2026cadrawer}. Closest to ours, Free2CAD~\cite{Free2CAD} and CADrawer~\cite{li2026cadrawer} parse strokes into sequences of CAD operations. While these methods reconstruct a single shape from a drawing, we generate multiple programs per part that can be independently sampled and recombined.

\paragraph*{CAD Program Generation.}

Prior work generates CAD programs from diverse inputs, including voxel grids~\cite{Sharma_2018_CVPR,ShapePrograms,LambournePrismatic}, point clouds~\cite{Wu2018, TaoInverseCSG, DeepCAD, Point2CAD, GuoComplexGen2022, sharma2020parsenet, Rukhovich2025CADRecode}, and boundary representations~\cite{ZoneGraphs}, among others~\cite{Chereddy2025SketchDNN,Wang2024VQCAD}. More recent work uses language and vision-language models for CAD program synthesis. Early approaches train or fine-tune models on CAD sequences~\cite{Xu2025CADMLLM,Rukhovich2025CADRecode,Khan2024VLMforCADFeatures,Li2025CADLlama,Khan2024CADSIGNet,Wang2025CADFusion,Doris2025CADCoder,Yuan2024OpenECAD,Zhang2025FlexCAD}, but are limited by the scale and operation coverage of available CAD datasets. Recent methods instead place off-the-shelf models in agentic loops, where they generate CAD code, inspect execution results and rendered views, and revise their output over multiple rounds~\cite{Alrashedy2024CADCodeVerify,Mallis2025CADAssistant,Ocker2025IdeaToCAD,Fan2025CADDesigner,Barkley2026CADSmith,ZeroToCAD2026}. We build on this training-free paradigm, but generate independently sampled programs for each part rather than a single CAD program. We reconcile their connecting geometry with a deterministic rewrite, making parts interchangeable across samples.

\paragraph*{Part-based generation and composition.}

Representing shapes as structured arrangements of parts is common in shape programs and procedural modeling, where programs encode both the parts and their relationships~\cite{mueller2006procedural,jones2020shapeassembly,jones2021shapemod,jones2023shapecoder,li2024possibleimpossibles}. Parametric design systems similarly expose part-based design spaces for interactive exploration~\cite{schulz2017interactive,Kazi2017,Matejka2018DreamLens,zhao2019carpentry}. We build on this idea, using CAD programs to represent and vary independently generated parts.

Recent systems also generate or compose shapes part by part. Compos3D~\cite{faruqi2026compos3d} synthesizes a coherent mesh from a user-arranged composition of parts. Recent vehicle-design systems combine parts across multiple views or sketch and generate parts individually~\cite{LeeCarExterior2025,YoonCarInterior2026}. ArtiCAD~\cite{ArtiCAD2026} generates articulated CAD assemblies from text or images, but does not address interchangeability across independently sampled variants. We target this problem directly, using sketch-derived relations to guide part generation and a deterministic rewrite to make variants interchangeable.

%% file: sections/3_Overview.tex
\section{Overview}
\label{sec:overview}

\begin{figure}[h!]
  \includegraphics[width=\linewidth]{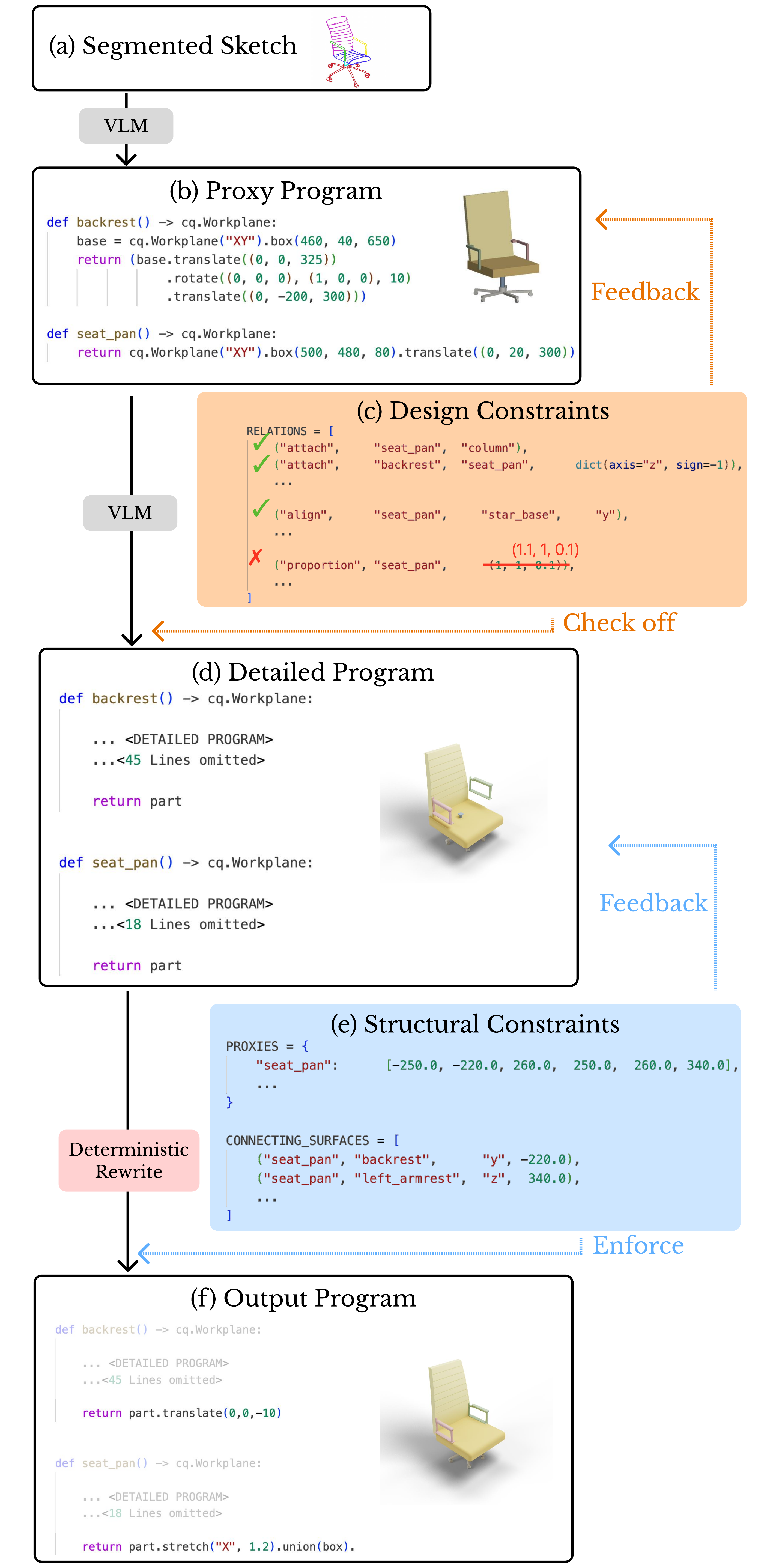}
\caption{Programs and constraints at each stage: (a) input; (b) proxy program; (c) design constraint feedback; (d) detailed program; (e) structural constraints (proxies and connecting surfaces); (f) rewrite, appending a wrapper while preserving the original code (gray). Parts are shown together for illustration; detail generation runs independently per part using its proxy program and corresponding sketch region.}
\label{fig:prog_change}
\end{figure}

Given a concept sketch, Composable CAD generates several CAD program variants for each component (Figure~\ref{fig:pipeline}). We give a high-level overview of the pipeline here; Appendices~\ref{appendix:sketch_segmentation}--\ref{appendix:sampling} provide full details.

We first plan the object as a coarse proxy, using geometric relations recovered from the sketch to establish how its components are sized, aligned, and attached. The proxy then provides the context for generating each component independently and defines the interfaces where neighboring components should meet. Finally, a deterministic rewrite seats each generated component against these interfaces. The rewritten components can be freely recombined: they do not overlap, meet exactly at shared interfaces, and form a single connected solid. A material budget limits how much the rewrite may alter the sampled geometry (Section~\ref{sec:program_rewrite}).

To ground this workflow in practice, we conducted a formative study with industrial designers prior to development (Appendix~\ref{appendix:formative_study}). Designers described an early phase of loose sketching followed by a 3D realization phase, where they work part by part and may divide components across collaborators. They also described producing useful design variants as time-consuming, particularly when variants require adding or reworking detail. Our system targets this later phase, supporting part-based exploration while keeping independently generated components mutually compatible~\cite{buxton2007sketching}.

We formulate composability through two types of constraints: \textit{design constraints}, set by the designer's sketch, and \textit{structural constraints}, which guarantee valid composition. Design constraints capture the sketch's intent: which components attach, how they align, and their relative proportions. Because these relations are recovered from an imprecise 2D drawing, we treat them as guidance rather than enforce them exactly. Structural constraints specify the exact geometry required for components to compose without gaps or volumetric overlap. We use the design constraints to construct the shared proxy, then derive the structural constraints from it.

Our pipeline has two stages (Figure~\ref{fig:pipeline}). The \textit{planning stage} uses design constraints from the sketch to construct a shared proxy and derive its structural constraints. The \textit{detail generation stage} then elaborates each component independently from the proxy, followed by a deterministic rewrite that enforces the structural constraints. Figure~\ref{fig:prog_change} shows these programs and constraints.

\subsection{Planning}

The planning stage reasons over the whole object once, establishing how its components belong together before any of them is elaborated. We first segment the sketch into semantic components (Appendix~\ref{appendix:sketch_segmentation}) and recover four types of design constraints: attachment, alignment, proportion, and mirroring. We extract these relations geometrically (Appendix~\ref{appendix:geo_relations}) rather than asking the VLM to infer them, since VLMs remain unreliable at quantitative spatial reasoning~\cite{liao2024qspatial, chen2024spatialvlm}. We detect stroke junctions, which identify attachments; junction-incident strokes then vote for the object's coordinate axes, and projecting component regions onto these axes gives alignment and proportion. Mirroring comes from the segmentation.

The design constraints guide the VLM in constructing a coarse proxy of the object, with one simple volume per component (Appendices~\ref{appendix:constraints_as_functions}-\ref{appendix:proxy_generation}). We compile each relation into a function that checks whether it holds on the generated proxy. Figure~\ref{fig:prog_change} shows this process for the chair: the backrest, for example, has constraints on its proportion, attachment to the seat pan, and alignment with neighboring components. We execute these checks and return any violations to the VLM for one round of refinement (Figure~\ref{fig:prog_change}c). We treat these relations as guidance rather than requirements: both their recovery and the underlying freehand sketch are approximate, so the VLM may overrule a relation that conflicts with the shape it is building. Mirroring is exact because we generate one member of a mirrored pair and reflect it for the other.

Once the proxy is fixed, we use it to determine exactly where neighboring components should meet. We represent each component by its axis-aligned proxy box and trim overlapping boxes so their interiors are disjoint. Where two attached boxes meet, their shared planar boundary defines a connecting surface (Figure~\ref{fig:prog_change}e). The resulting proxy boxes and connecting surfaces form the structural constraints: each component must remain within its proxy and meet its neighbors at the prescribed surfaces.

\subsection{Independent Detail Generation}
Once planning is complete, we elaborate each component independently and in parallel; see Appendix~\ref{appendix:sampling} for full details. In the detail generation stage, we provide the VLM with only the component's proxy program and corresponding sketch region. Rather than generating a new program from scratch, the VLM edits the proxy program, which represents the component as a simple volume, and elaborates it into detailed geometry. The proxy carries the global information needed to place the component, while the sketch region provides the visual information needed to elaborate it. We draw several variants for each component from independent VLM calls.

We can generate components independently and in parallel because each component does not need to see the whole object, only enough context to know where it belongs. Giving the VLM the whole object instead introduces irrelevant context that can degrade generation~\cite{shi2023distracted,liu2025logomotion}; in our evaluation, focused per-component generation produces more detailed programs (Section~\ref{subsec:eval_qualitative}).

While the proxy guides placement, it does not guarantee that the detailed geometry still reaches its connecting surfaces. We therefore apply a deterministic rewrite (Section~\ref{sec:program_rewrite}) to seat each sampled component against these surfaces before adding it to the design space.

%% file: sections/4_DeterministicAlgorithm.tex
\section{Deterministic Rewrite}
\label{sec:program_rewrite}

\begin{figure}[t!]
  \includegraphics[width=\linewidth]{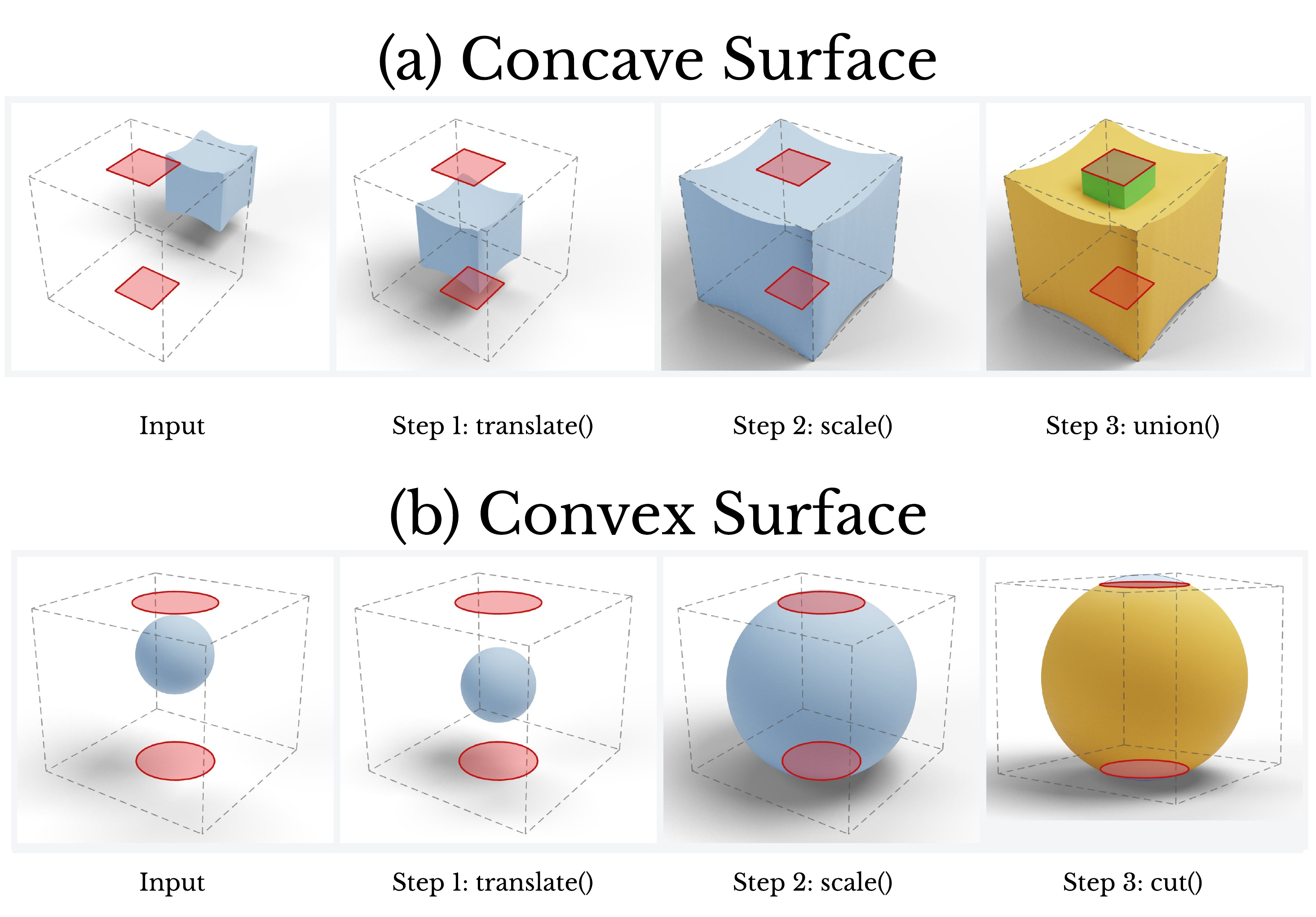}
\caption{The three rewrite steps on a concave part (a) and a convex one (b). The gray dashed region is the part's proxy and the red translucent regions are its connecting surfaces, both fixed by the planning stage. In (a),
the green region shows only the material introduced to establish the prescribed interface, and
in (b) the red region marks material cut from a face that protrudes past it. The material budget limits the total geometry the rewrite may change.}
\label{fig:det_rewrite}
\end{figure}

\begin{algorithm}[tbp]
\caption{Deterministic rewrite of a part}
\label{alg:rewrite}
\KwIn{solid $V_c$, proxy $B_c$, connecting surfaces
  $S_c = \{s_1,\dots,s_n\}$ on planes $\{\pi_1,\dots,\pi_n\}$}
\KwOut{rewritten solid $V'_c$}
\BlankLine
\tcp{Step 1: translate to the proxy center}
$V_c \leftarrow V_c + \bigl(\mathrm{center}(B_c)
  - \mathrm{center}(V_c)\bigr)$\;
\BlankLine
\tcp{Step 2: scale per axis about $\mathrm{center}(B_c)$}
\ForEach{axis $a \in \{x,y,z\}$}{
  \eIf{a connecting surface lies on a face normal to $a$}{
    $\lambda_a \leftarrow$ smallest factor reaching every such surface:
      contact with $\pi_i$ where $V_c$ is concave, coverage of $s_i$
      where it is convex\;
  }{
    $\lambda_a \leftarrow \min(1, \text{largest factor keeping } V_c
      \text{ inside } B_c \text{ along } a)$\;
  }
}
$V_c \leftarrow \mathrm{scale}(V_c, \lambda_x, \lambda_y, \lambda_z)$
  about $\mathrm{center}(B_c)$\;
\BlankLine
\tcp{Step 3: form a flat face at each connecting surface}

\ForEach{$s_i \in S_c$}{
  reclassify $V_c$ at $s_i$\;
  \eIf{convex}{
    cut away the material of $V_c$ lying beyond $\pi_i$ over a
      neighborhood of $s_i$\;
  }{
    union $V_c$ with the solid of cross section $s_i$ extruded
      inward from $\pi_i$ until it meets the part\;
  }
}
\Return $V_c$\;
\end{algorithm}

We now describe our deterministic rewrite algorithm for seating independently generated components against the structural constraints defined during planning.

\subsection{Seating a Component}

Our deterministic rewrite enforces the structural constraints while preserving the generated geometry wherever possible. It seats each component inside its proxy and against its connecting surfaces using three increasingly geometry-changing operations (Figure~\ref{fig:det_rewrite}).
We say a component is seated when it remains within its proxy and covers each connecting surface without extending beyond its interface plane. We first translate the component to the center of its proxy, then scale it to reach its connecting surfaces while remaining within the proxy. Finally, we form a flat face at each interface: if the generated surface falls short of the connecting surface, we union material to fill the gap; if it protrudes beyond it, we cut the excess. Figure~\ref{fig:det_rewrite} shows both cases. These operations modify only the residual geometry needed to establish the interface; we bound the total modification with the material budget described in Section~\ref{subsec:budget}.

We perform this step geometrically rather than ask the VLM to repair its program because connecting surfaces are often implicit in CAD code. An extrusion, for example, creates faces with no explicit handle in the program; their positions are known only after execution. Placing such a face at a prescribed interface therefore requires reasoning about geometry that the program does not explicitly represent. Predicting program outputs remains unreliable even for short functions~\cite{gu2024cruxeval}, and quantitative spatial reasoning is itself difficult for VLMs~\cite{liao2024qspatial, chen2024spatialvlm}. We instead operate on the executed geometry and append deterministic CadQuery operations to the sampled program.

\subsection{Rewrite Algorithm}
Let \(P_c\) be the sampled program for component \(c\), \(V_c\) its executed solid, \(B_c\) its proxy, and \(S_c=\{s_1,\ldots,s_n\}\) its connecting surfaces. Algorithm~\ref{alg:rewrite} rewrites \(P_c\) by appending translation, scaling, union, and cutting operations. We append rather than modify the sampled construction to preserve as much of the VLM-generated geometry as possible.

The operations are applied from least to most geometry-changing. First, we translate the component to the center of its proxy, preserving its geometry exactly. Second, we scale each axis until the component reaches its connecting surfaces while remaining within its proxy; scaling changes dimensions but preserves its features and their arrangement. Third, we classify the generated surface at each interface as concave or convex. A concave surface falls short of the interface, so we union an extrusion of the connecting surface inward until it reaches the component. A convex surface protrudes beyond it, so we cut away the excess. Appendix~\ref{appendix:suface_analysis_counter_case} explains why this choice preserves more of the sampled geometry.

\begin{figure*}[t!]
\centering
\includegraphics[width=0.74\linewidth, height=4.35cm]{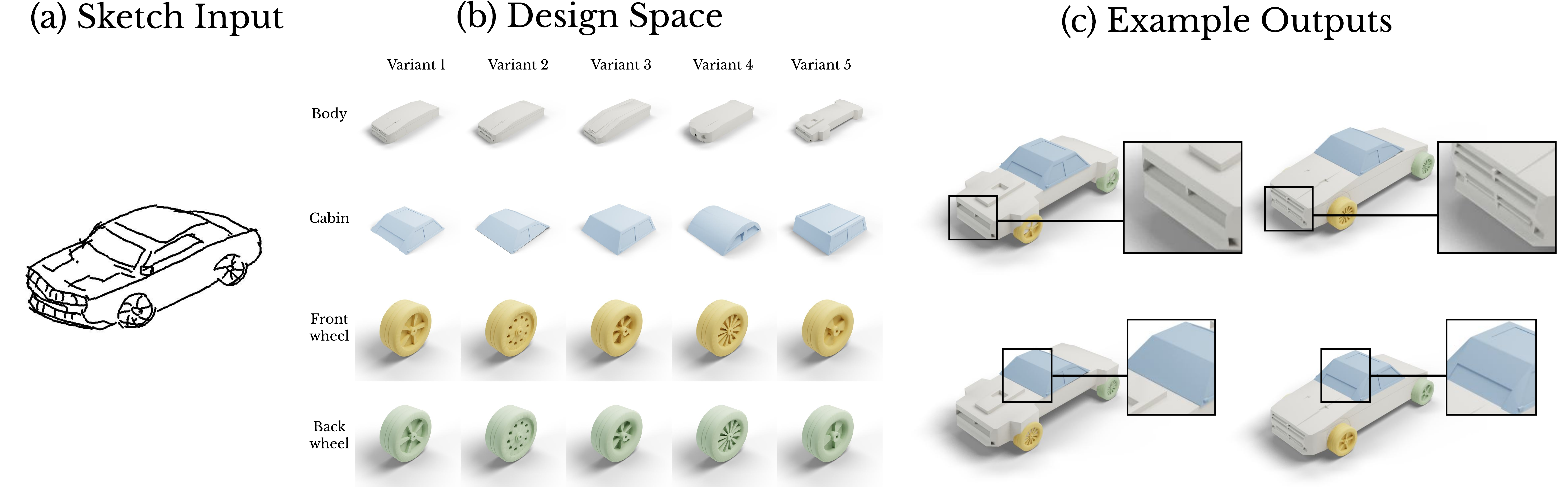}
\vspace{-5pt}
\caption{Design space for a sports car sketch. Insets highlight the components that differ between the shown outputs.}
\label{fig:result_car_2}
\end{figure*}

\begin{figure*}[t!]
\centering
\includegraphics[width=0.74\linewidth, height=3.515cm]{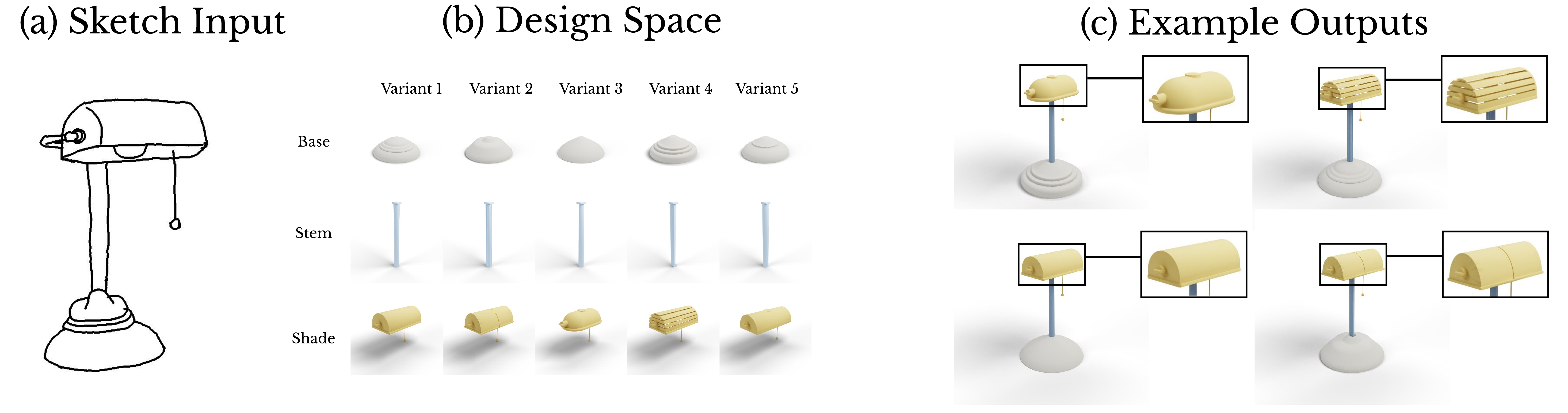}
\vspace{-5pt}
\caption{Design space for a banker's lamp, segmented into three components. Insets highlight the components that differ between the shown outputs.}
\label{fig:result_lamp}
\end{figure*}

\begin{figure*}[t!]
\includegraphics[width=\linewidth, height=3.2cm]{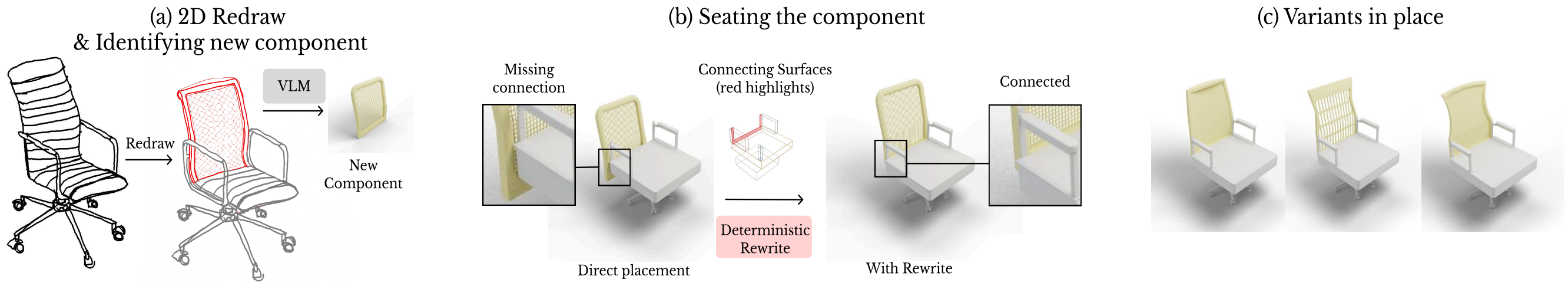}
\vspace{-5pt}
\caption{Editing a component in 2D. The designer redraws the backrest, and the VLM samples a new component from the sketch (a). Placed directly it misses its neighbors (b, left inset); assigning it the connecting surfaces the old backrest left (red) and running the deterministic rewrite seats it flush (b, right inset). Sampling yields several backrests, each seating into the same fixed chair (c). Only the backrest changes; the seat, base, and armrests are identical throughout.}
\label{fig:2D_Edit}
\end{figure*}

\begin{figure*}[t!]
\centering
\includegraphics[width=0.65\linewidth]{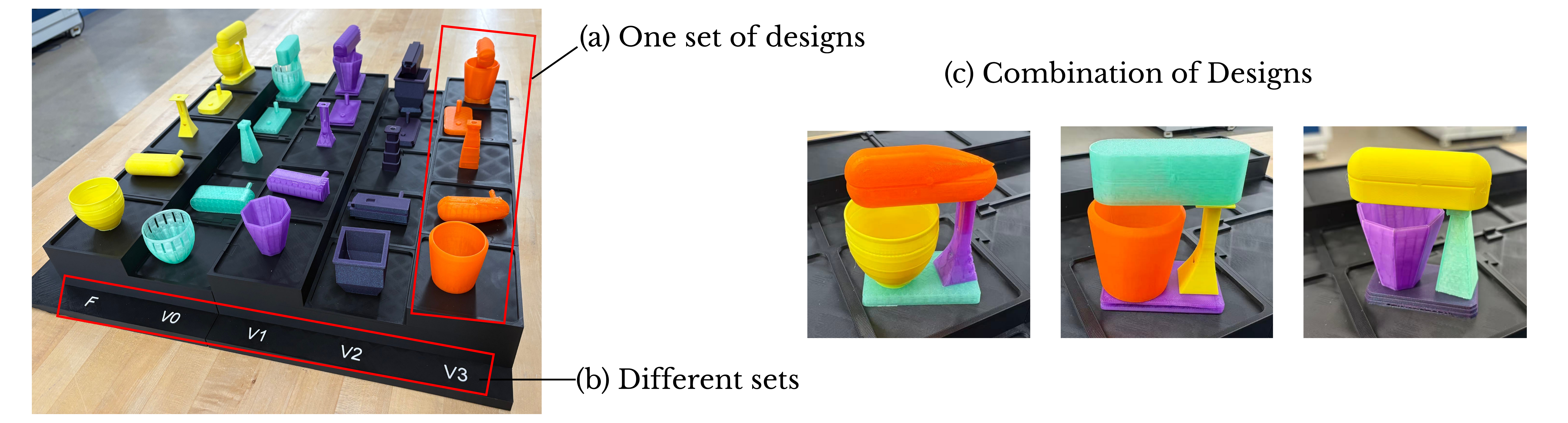}
\vspace{-5pt}
\caption{Fabricating a design space. We printed every sample of every component of the stand mixer, one column per sample index (b), so that a column read top to bottom gives one complete set of parts (a). Every part was generated independently, and so was every joint, yet parts can be taken from any column and assembled: the four mixers in (c) are built from different combinations of the printed parts, each meeting flush at its connecting surfaces.}
\label{fig:Fabrication}
\end{figure*}

\subsection{Composability Guarantee}
Because every variant is rewritten against the same proxy and connecting surfaces, the rewrite can operate on each component independently while ensuring that the resulting variants remain interchangeable.

\begin{proposition}.
Let the component proxies have pairwise disjoint interiors, with each attached pair sharing one planar connecting surface. Under the sample conditions of Appendix~\ref{appendix:proof}, if Algorithm~\ref{alg:rewrite} completes within the material budget for every component, then any selection of one rewritten sample per component has pairwise disjoint interiors, every attached pair meets exactly along its shared connecting surface, and their union is one connected solid.
\end{proposition}

The sample conditions require nonempty, connected material within each proxy and that the generated geometry can be seated at each interface by the corresponding cut or union; Appendix~\ref{appendix:proof} states them formally. Intuitively, keeping every component inside its proxy prevents volumetric overlap, while neighboring components cover the same connecting surface without extending beyond it. Since the attachment graph is connected, these shared interfaces make the assembled object connected.

The guarantee does not bound how much the rewrite must change a sampled component. A highly distorted interface may require a large cut or union, which would defeat the purpose of preserving the sampled geometry. We therefore cap the volume changed by scaling, cutting, and union at 10\% of the shape's total volume, and report failure when a component cannot be seated within this budget. Section~\ref{subsec:budget} evaluates this tradeoff.

%% file: sections/6_Applications.tex
\section{Applications}
\label{sec:Applications}

Composable CAD supports three applications: exploring the part-based design space, editing individual components, and fabricating interchangeable parts.

\subsection{Design-Space Exploration}
\label{subsec:design_space}

Our system exposes a part-based design space for exploring variations of the input sketch. The designer browses alternatives for each component and composes a design by selecting one of each; because the components are interchangeable, any selection can be freely recombined. We generate five variants per component independently and in parallel from the same proxy, varying the sampling temperature from 0.1 to 0.5 to produce a range of designs (Appendix~\ref{appendix:sampling}).

We show results on sketches from OpenSketch~\cite{gryaditskaya2019opensketch}, DifferSketching~\cite{xiao2022differsketching}, SlowSketch~\cite{bhunia2020pixelor}, and AmateurSketch~\cite{qi2021amateursketch}, spanning a range of drawing styles and levels of detail. For each, we show the input sketch, generated design space, and several recombined outputs (Figures~\ref{fig:result_car_2}, \ref{fig:result_lamp},  \ref{fig:result_car_1}, \ref{fig:result_mixer}, \ref{fig:result_seat}, \ref{fig:result_computer}, \ref{fig:result_vacuum}, \ref{fig:result_flan}). Insets highlight components that differ across outputs and remain compatible when recombined.

\subsection{Editing}
\label{subsec:editing}

Our structural constraints make component-level editing natural: removing a component leaves its proxy and connecting surfaces fixed, so any replacement inherits exactly the constraints of the original. The deterministic rewrite (Section~\ref{sec:program_rewrite}) seats the replacement against these interfaces, while every other component remains unchanged.

We demonstrate three ways to provide a replacement: redrawing the component as a sketch (Figure~\ref{fig:2D_Edit}), borrowing a finished component from another design (Figure~\ref{fig:3D_Edit}), or describing its replacement in text (Figure~\ref{fig:text_edit}). In each case, only the selected component changes. The replacement remains compatible with the full design space, so it can be recombined with variants of every other component and exploration can continue from the edited design.

\subsection{Fabrication}
\label{subsec:Fabrication}

Composable CAD also supports physical assembly: the flat, congruent connecting surfaces produced by our system provide natural interfaces for joining components. We demonstrate this by 3D printing an entire design space---every sample of every component---and assembling designs from components across different samples (Figure~\ref{fig:Fabrication}). Because the components meet at the same connecting surfaces, the recombination available in the digital design space carries over to the physical parts.

The joints are independent as well. For each connecting surface, we place a peg on one component and a socket on the other, with no coordination between their generation. Since both are derived from the same connecting surface, they are congruent by construction (Appendix~\ref{appendix:Joints}). See the supplemental video for additional examples.

%% file: sections/7_Evaluations.tex
\section{Evaluation}
\label{sec:evaluation}

We evaluate Composable CAD on the properties our method is designed to provide: preserving the sketch's design constraints, maintaining structural validity under part recombination, and producing detailed CAD programs, together with the cost of generation. We compare against VLM-only baselines across a corpus of objects and sketch styles, and complement them with qualitative results and two user studies.

\subsection{Baselines}

We compare against two VLM-only baselines and one ablation of our method. The first tests direct sketch-to-CAD generation with an agentic VLM; the second gives the VLM the same coarse-to-fine generation strategy as our method, but without our geometric constraints or deterministic rewrite. Finally, we compare our pipeline with and without the deterministic rewrite to isolate its effect on composability. Prompts for the VLM baselines are given in Appendix~\ref{appendix:setup_prompts}.

\textbf{S1: VLM one pass.}
We use this baseline to represent a practical single-pass setting. Given the sketch, the VLM segments the object and generates one CadQuery program with one function per component. We use high thinking with a five-minute budget.

\textbf{S2: VLM two passes.}
This baseline tests whether coarse-to-fine generation alone is sufficient. The VLM first generates a whole-object proxy, then elaborates each component with the full proxy program as context. It performs its own segmentation and uses no geometric constraints or deterministic rewrite. We use high thinking with no time limit, allowing the VLM to complete the more time-consuming two-stage generation.

\textbf{S3: Ours without rewrite.}
We run our full pipeline but disable the deterministic rewrite (Section~\ref{sec:program_rewrite}). Sketch-derived relations still guide proxy generation, and components are independently elaborated from their proxies, but they are not reconciled at their connecting surfaces.

\textbf{Ours.}
Our complete pipeline, using the same VLM with medium thinking.

\subsection{Evaluation Protocol}
\label{subsec:evaluation_setup}

We evaluate on $33$ objects randomly selected from the sketch datasets used in Section~\ref{subsec:design_space}, spanning different object types, drawing styles, and levels of detail.
Each method generates five program variants per object. We evaluate these programs both as generated and after recombination. \textit{As generated} evaluates each program directly, with all components coming from the same variant, giving $33\times5=165$ shapes per method. \textit{Recombined} forms 25 additional shapes per object by randomly selecting one component from across the five variants, giving $33\times25=825$ shapes per method. This tests whether components generated in different samples remain interchangeable. Shapes are normalized to a $10\times10\times10$ cm bounding box before geometric measurement. Non-executable outputs are rare, since each setup executes its generated programs and uses the VLM to repair compilation errors.

\begin{table}[t]
\centering
\small
\begin{tabular}{l c c c c}
\toprule
& $S_1$ & $S_2$ & $S_3$ & Ours \\
\midrule
Proportion $\uparrow$ & 80.8\% & 74.6\% & \textbf{91.2\%} & 89.4\% \\
Alignment $\uparrow$  & 84.7\% & 62.5\% & \textbf{88.8\%} & 86.9\% \\
Mirroring $\uparrow$  & 93.0\% & 82.3\% & \textbf{100\%} & \textbf{100\%} \\
\bottomrule
\end{tabular}
\caption{Fraction of design-constraint relations satisfied, for the setups of Section~\ref{subsec:evaluation_setup}, checked with the relation functions described in the text. Bold marks the best per row. $S_3$, our pipeline without the rewrite, scores highest on proportion and alignment; Ours is slightly below it because the rewrite moves geometry after the relations have been checked. Both reach 100\% on mirroring by construction, since mirrored parts are generated once and reflected.}
\label{tab:design}
\end{table}

\begin{table*}[t]
\centering
\small
\begin{tabular}{l c c c c c c c c}
\toprule
& \multicolumn{2}{c}{$S_1$} & \multicolumn{2}{c}{$S_2$} & \multicolumn{2}{c}{$S_3$} & \multicolumn{2}{c}{Ours} \\
\cmidrule(lr){2-3} \cmidrule(lr){4-5} \cmidrule(lr){6-7} \cmidrule(lr){8-9}
& generated & recombined & generated & recombined & generated & recombined & generated & recombined \\
\midrule
\# Shapes                  & 165 & 825 & 165 & 825 & 165 & 825 & 165 & 825 \\
\midrule
No Overlap $\uparrow$      & 92.1\% & 23.9\% & 81.2\% & 79.2\% & 80.0\% & 78.9\% & \textbf{95.8\%} & \textbf{94.1\%} \\
Attached $\uparrow$        & 87.3\% & 27.6\% & 80.0\% & 81.5\% & 81.8\% & 82.7\% & \textbf{95.8\%} & \textbf{94.9\%} \\
Valid Interface $\uparrow$ & 73.9\% & 9.9\%  & 55.8\% & 54.5\% & 53.9\% & 55.2\% & \textbf{93.3\%} & \textbf{91.5\%} \\
\midrule
All Succeed $\uparrow$     & 73.9\% & 9.9\%  & 55.8\% & 54.5\% & 53.9\% & 55.2\% & \textbf{93.3\%} & \textbf{91.5\%} \\
\bottomrule
\end{tabular}
\caption{Structural validity scored on variants as generated and on recombined selections. Each score is the fraction of shapes satisfying it; All Succeed is the fraction satisfying all three. Our rewrite operates under a 10\% material budget. Bold marks the best per row.}
\label{tab:compat}
\end{table*}

\begin{table}[t]
\centering
\small
\begin{tabular}{l c c c c}
\toprule
& 1\% & 5\% & 10\% & 20\% \\
\midrule
No Overlap $\uparrow$      & 81.2\% & 93.7\% & 94.1\% & 99.2\% \\
Attached $\uparrow$        & 82.1\% & 91.0\% & 94.9\% & 96.2\% \\
Valid Interface $\uparrow$ & 55.0\% & 86.8\% & 91.5\% & 94.4\% \\
\midrule
All Succeed $\uparrow$     & 55.0\% & 86.8\% & 91.5\% & 94.4\% \\
\bottomrule
\end{tabular}
\caption{Structural validity as the material budget varies, scored on recombined selections (825 shapes). The budget caps the volume the rewrite may add or remove as a fraction of the shape's total volume. We use 10\% throughout the paper.}
\label{tab:budget}
\end{table}

\begin{table}[t]
\centering
\small
\begin{tabular}{l c c c c}
\toprule
& $S_1$ & $S_2$ & $S_3$ & Ours \\
\midrule
CAD operations  & 53.2 & 85.6 & 113.7 & \textbf{127.5} \\
Lines of code   & 77.9 & 104.5 & \textbf{274.8} & \textbf{274.8} \\
\bottomrule
\end{tabular}
\caption{Program complexity for the setups of Section~\ref{subsec:evaluation_setup}, over 33 objects with five runs each, averaged over all runs. Bold marks the best per row. $S_3$ and Ours share a line count because the rewrite appends its operations to the returned part rather than adding lines of its own, so it raises the operation count while leaving the program length unchanged.}
\label{tab:detail}
\end{table}

\begin{table}[t]
\centering
\small
\begin{tabular}{l c c c}
\toprule
& $S_3$ & VLM repair & Ours \\
\midrule
\# Shapes                  & 825 & 825 & 825 \\
\midrule
No Overlap $\uparrow$      & 78.9\% & 91.2\% & \textbf{94.1\%} \\
Attached $\uparrow$        & 82.7\% & 89.5\% & \textbf{94.9\%} \\
Valid Interface $\uparrow$ & 55.2\% & 54.1\% & \textbf{91.5\%} \\
\midrule
All Succeed $\uparrow$     & 55.2\% & 54.1\% & \textbf{91.5\%} \\
\midrule
Repair time & -- & 3 min (cap) & Negligible \\
\bottomrule
\end{tabular}
\caption{Structural validity compared with using a VLM for repair, evaluated on the same 825 recombined selections. Each structural metric is the fraction of shapes satisfying it; All Succeed requires all three. Bold marks the best structural result per row. Our rewrite uses the standard 10\% material budget. The VLM time entry denotes its three-minute wall-time cap per shape.}
\label{tab:vlm_repair}
\end{table}

\begin{table}[t]
\centering
\small
\begin{tabular}{l c c c c}
\toprule
& $S_1$ & $S_2$ & $S_3$ & Ours \\
\midrule
Time (min)   & 5 (capped) & 9--44 & 5--11 & 5--11 \\
Tokens (mean) & 228k & 371k & 526k & 526k \\
Tokens (range)            & 131--337k & 216--546k & 349--688k & 349--688k \\
\bottomrule
\end{tabular}
\caption{Generation cost per object, over 33 objects with five runs each, using Gemini~3.1~Pro~\cite{gemini31pro2026}. Token usage is reported as mean and range. Generation time is reported as a range, except for $S_1$, for which we report the five-minute cap; $S_2$ runs uncapped. Costs vary with the number and complexity of parts. $S_3$ and Ours have the same token usage and reported time ranges, since the rewrite adds negligible runtime and consumes no tokens.}
\label{tab:timing}
\end{table}

\subsection{Quantitative Results}
\label{subsec:eval_quantitative}
We quantify how well each method preserves the sketch constraints, maintains composability, and generates detailed CAD programs.

\textbf{Design Constraints.}
To measure satisfaction of design constraints, we evaluate how well the generated shapes preserve the alignment, proportion, and mirroring relations recovered from the sketch. We reuse the executable relation functions of Appendix~\ref{appendix:constraints_as_functions}: proportion compares a component's $x{:}y{:}z$ extents with the ratio recovered from the sketch, alignment checks the specified component anchors along each axis, and mirroring checks that mirrored components match in extent. We omit attachment because physical contact is enforced by the structural constraints and evaluated separately below.

We report the fraction of extracted relations that pass in Table~\ref{tab:design}. Because $S_1$ and $S_2$ determine their own segmentation, we extract relations using each setup's own part list. Results show that our planning stage substantially preserves these relations. $S_3$ scores highest on proportion and alignment, while our complete method is slightly lower because the rewrite modifies geometry after the relations have been checked. Neither is expected to reach 100\%: design constraints are soft guidance, and the VLM may overrule a recovered relation when constructing the proxy. Both $S_3$ and ours satisfy mirroring by construction because mirrored components are generated once and reflected.

\begin{figure}[t!]
  \includegraphics[width=\linewidth]{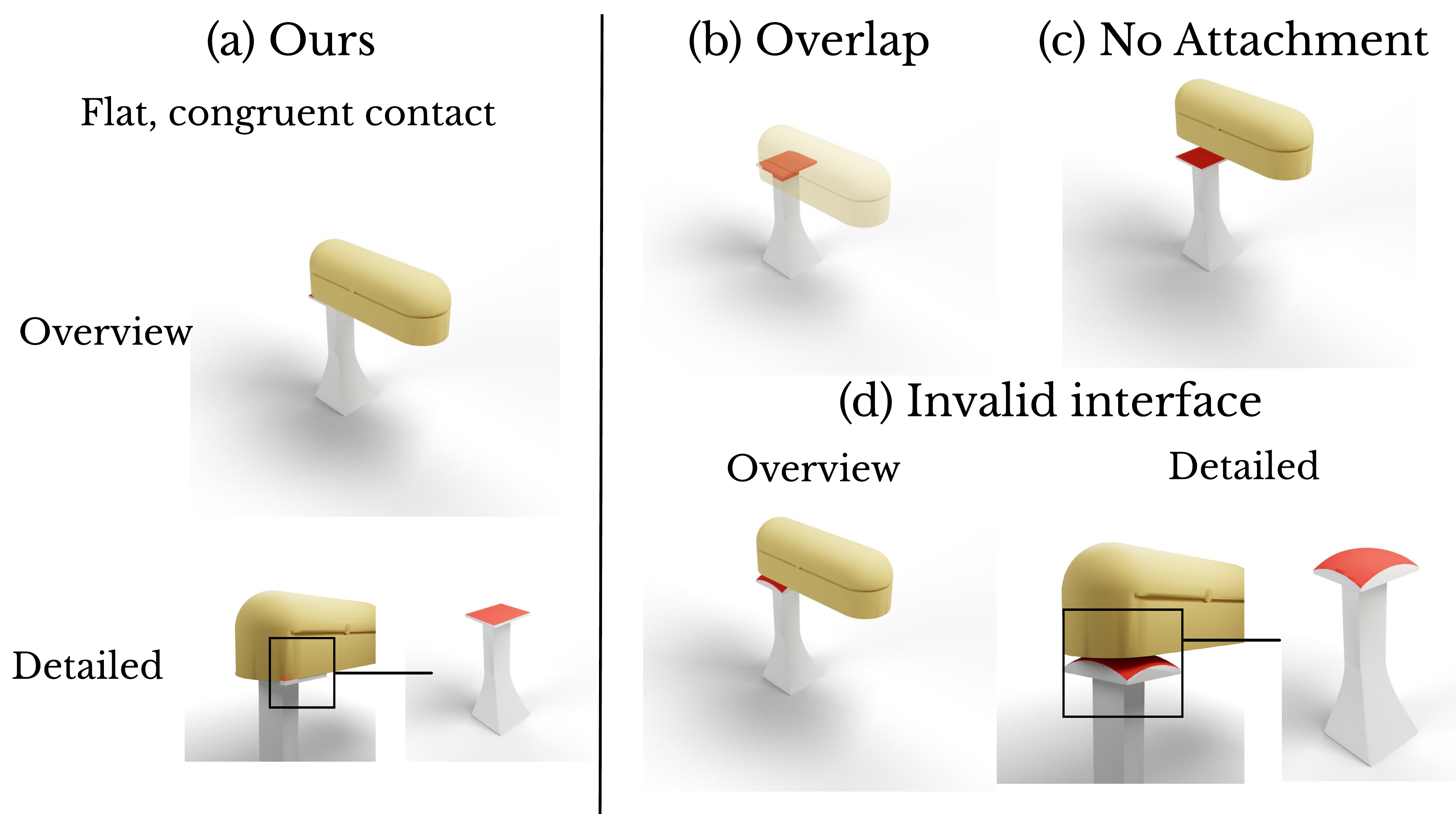}
\caption{Example failures without deterministic rewrite. The red region is the connecting surface between the motor head and the stand. Our rewrite seats the two so that they meet flush over the whole surface, and the isolated stand shows the flat face it produces (a). Without it, three failures occur: the parts interpenetrate (b), they do not touch at all (c), or they touch but not over a flat region, here a domed face that meets the head along a curve rather than a face (d).}
\label{fig:structure_failure_demo}
\end{figure}

\textbf{Structural Validity.}
We next evaluate whether independently generated components compose correctly, particularly after recombination. We measure three conditions. \textit{No overlap} requires that no two components intersect. \textit{Attached} requires every pair marked as attached by the sketch to have surfaces within 0.1 cm. \textit{Valid interface} requires each attached pair to meet over a flat contact region whose side exceeds $\max(0.3\text{ cm}, t)$, where $t$ is the component's cross-sectional thickness at the interface. Figure~\ref{fig:structure_failure_demo} illustrates these failure modes. For our method, the interface is the computed connecting surface; because the VLM baselines have no explicit interface, we take it as the region of one component's boundary within 0.1 cm of the other.

Table~\ref{tab:compat} reports these conditions both as generated and after recombination. Our method maintains similar structural validity in both settings: 93.3\% of generated shapes and 91.5\% of recombined shapes satisfy all three conditions. In contrast, the single-pass baseline drops from 73.9\% to 9.9\% after recombination, since components generated together may not remain compatible when mixed across samples. $S_2$ and $S_3$ likewise remain substantially below our method, showing the role of the deterministic rewrite in making independently sampled components interchangeable.

\textbf{Material Budget.}
\label{subsec:budget}
The material budget limits how much the rewrite may alter a component after detail generation. We measure rewrite cost as the volume changed by scaling, cutting, and union, and cap it at 10\% of the shape's total volume; translation does not contribute because it preserves geometry exactly. If any component cannot be seated within this budget, the composed shape is scored as failing all three structural conditions.

Table~\ref{tab:budget} varies the budget from 1\% to 20\%. A tight budget preserves more of the sampled geometry but leaves the rewrite too little room to establish some interfaces, while a larger budget improves structural validity at the cost of modifying more geometry. We use 10\% throughout the paper, at which 91.5\% of recombined shapes satisfy all three structural conditions.

\begin{figure}[t!]
  \includegraphics[width=\linewidth]{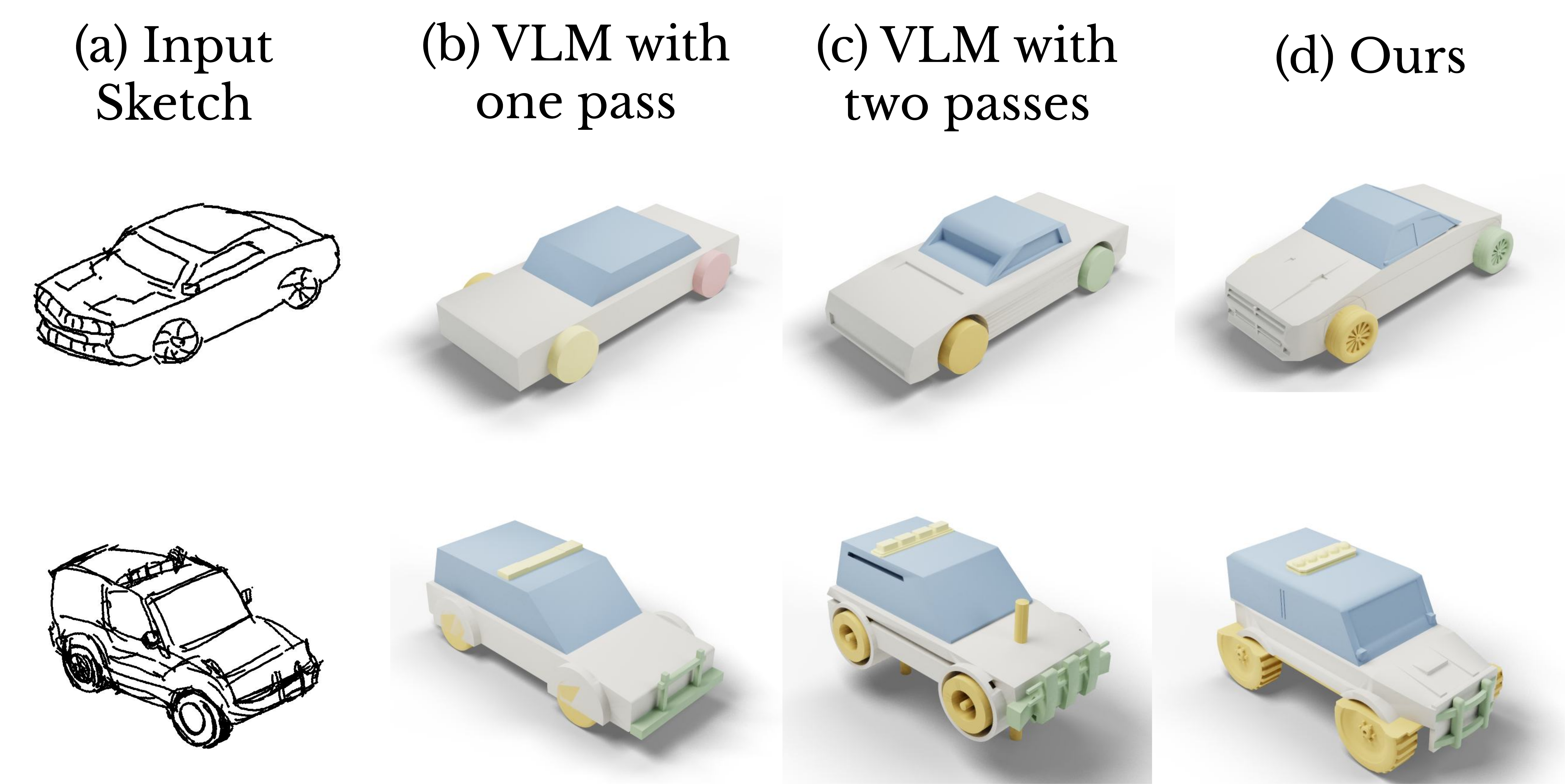}
\caption{Detail across generation strategies. Given the input sketch (a), a single pass produces parts that are little more than primitives: flat body slabs, plain cylindrical wheels, and no grille or bumper structure (b). Giving each part its own pass adds detail (c), and our pipeline adds more still (d), with spoked and treaded wheels, slatted grilles and bumpers, and panel creases on the body. Detail is one benefit of elaborating parts separately rather than generating the whole object at once.}
\label{fig:2Pass_more_details}
\end{figure}

\textbf{Program Detail.}
Component-wise generation produces substantially richer CAD programs. Table~\ref{tab:detail} reports number of CAD operations and lines of code across the four setups. The one-pass baseline averages 53.2 CAD operations, compared with 85.6 for the two-pass VLM baseline and 127.5 for our complete method. S3 and ours have the same line count because the rewrite appends operations to the returned component rather than adding source lines, increasing the operation count without changing program length. Figure~\ref{fig:2Pass_more_details} shows the corresponding qualitative differences in generated geometry.

\textbf{VLM-Based Repair}
Our deterministic rewrite raises a natural alternative: can the VLM itself repair independently generated components? We test this on the same 825 recombined S3 shapes used above. Gemini 3.1 Pro receives the complete CadQuery program and may repeatedly execute and revise it for up to three minutes to remove gaps and overlaps, produce flat interfaces, and preserve component detail. Appendix~\ref{appendix:vlm_repair} includes the full prompt.

Table~\ref{tab:vlm_repair} shows that VLM repair improves overlap and attachment, but not valid interfaces: the fraction satisfying all structural conditions changes from 55.2\% before repair to 54.1\% afterward. Our deterministic rewrite reaches 91.5\% with negligible additional runtime. This advantage likely comes from operating explicitly in 3D: the VLM must reason about interfaces indirectly through CAD code, whereas our rewrite measures and modifies the executed geometry directly, making it better suited to enforcing the precise interfaces required for composability.

\textbf{Generation Cost}
Finally, Table~\ref{tab:timing} reports generation time and token usage. The one-pass baseline is fastest and uses the fewest tokens, but produces substantially less detailed programs. Generating components independently in parallel makes $S_3$ and ours faster than sequential two-pass generation. Our pipeline uses more tokens than the VLM-only baselines because it also reasons over design-constraint feedback. The deterministic rewrite itself adds negligible runtime and consumes no VLM tokens.

\begin{figure}[t!]
  \includegraphics[width=\linewidth]{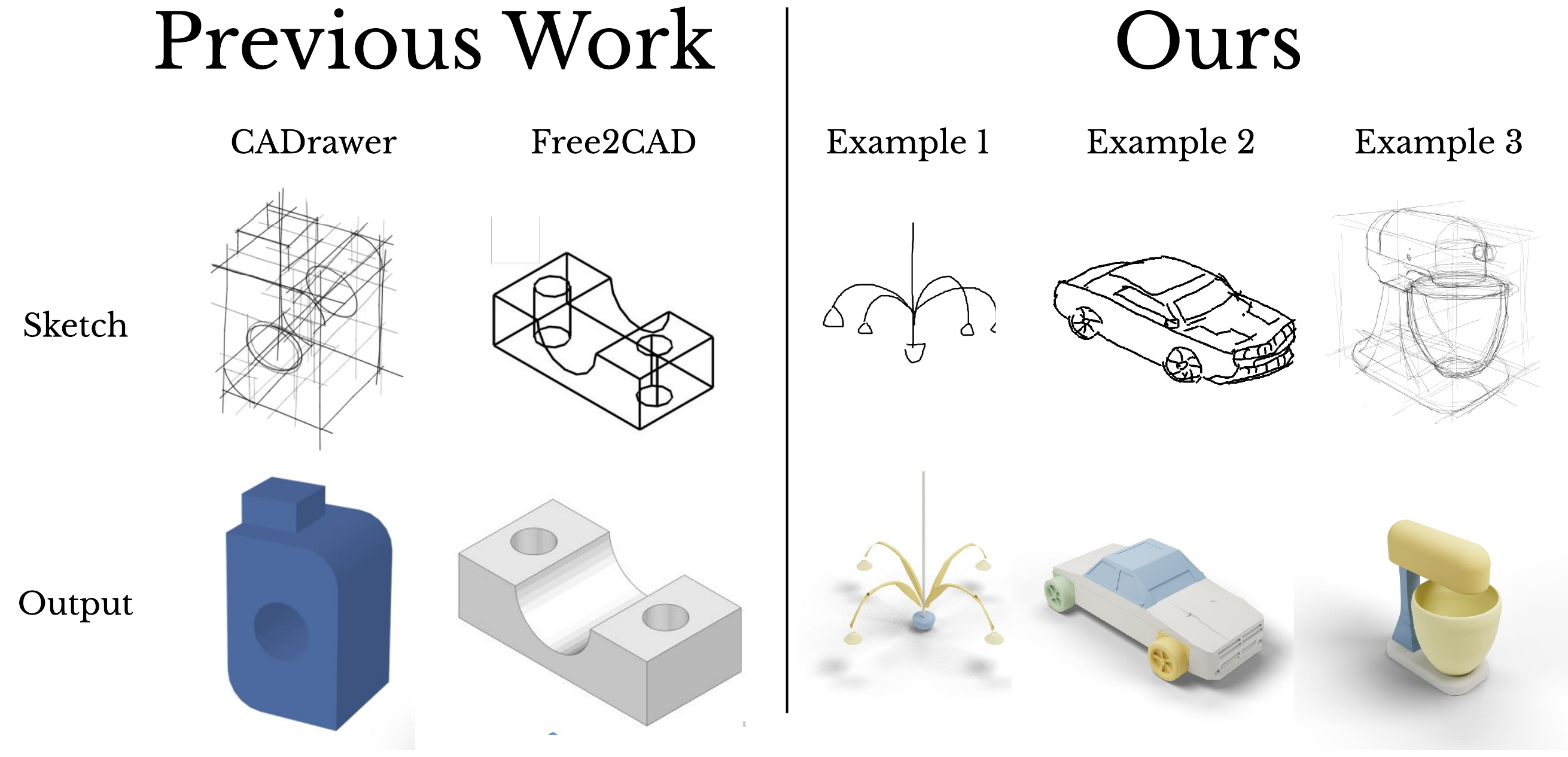}
\caption{Scope compared to prior sketch-to-CAD systems. CADrawer~\cite{li2026cadrawer} and Free2CAD~\cite{Free2CAD} take clean line drawings of a single part and return one reconstruction. Our inputs span a wide range of drawing styles and complexity, and our outputs are multi-part objects with substantially longer programs, more than 100 CAD operations for the car.}
\label{fig:sketch_handling}
\end{figure}

\subsection{Qualitative Results}
\label{subsec:eval_qualitative}
We show qualitative results of our method compared with baselines and with prior systems.
Figure~\ref{fig:2Pass_more_details} illustrates the benefit of component-wise generation. A single VLM pass captures the coarse shape but often omits component detail, such as wheel structure, grilles, bumpers, and body features. Giving components their own generation produces substantially richer geometry, and our per-component generation adds further detail.

This added detail comes with a challenge: independently generated components may not fit together. Figure~\ref{fig:structure_failure_demo} shows the three failure modes we observe: components may overlap, fail to touch, or touch over an invalid interface. Our deterministic rewrite resolves these failures by seating components against their shared connecting surfaces, producing flat, congruent contact while retaining their detail.

Beyond these differences in detail and composability, our system also targets a broader setting than prior sketch-to-CAD systems (Figure~\ref{fig:sketch_handling}). CADrawer~\cite{li2026cadrawer} and Free2CAD~\cite{Free2CAD} reconstruct a single part from a clean line drawing, whereas our inputs span a broader range of drawing styles and complexity and our outputs are multi-part design spaces rather than single reconstructions. CADrawer reports quality degrading markedly beyond roughly 15 CAD operations; our programs can be substantially longer, with around 100 operations for the car in Figure~\ref{fig:sketch_handling}.

\begin{table}[t]
\centering
\small
\begin{tabular}{l c c}
\toprule
& Ours chosen & $p$ \\
\midrule
Detail     & 77.8\% & $3.3\times10^{-14}$ \\
Proportion & 66.7\% & $9.1\times10^{-6}$ \\
Fidelity   & 70.0\% & $8.2\times10^{-8}$ \\
Overall    & 75.6\% & $3.9\times10^{-12}$ \\
\bottomrule
\end{tabular}
\caption{Human preference pairwise comparison between ours and the single-pass baseline. Each row is the percentage of choices in which our shape was chosen; the baseline takes the remainder, since participants could not abstain. The $p$-values are from an exact two-sided binomial test against chance. Detail asks which shape models more of the drawing's fine detail, proportion which has the parts at the right size and in the right place, fidelity which better depicts the sketched object in a buildable form, and overall which is the better result all things considered.}
\label{tab:user}
\end{table}

\subsection{Perceptual Preference Study}
\label{subsec:perceptual_study}
We next evaluate the perceptual quality of our generated outputs relative to direct VLM generation. We compare our method with the single-pass VLM baseline in a preference study with 12 participants, each evaluating 15 randomly selected sketches. For each sketch, participants saw the two outputs side by side in randomized, unlabeled order and indicated their preference on detail, proportion, fidelity to the input sketch, and overall quality. Appendix~\ref{appendix:User_protocol_1} gives the full protocol.

Participants preferred our outputs on all four criteria (Table~\ref{tab:user}): 77.8\% for detail, 66.7\% for proportion, 70.0\% for fidelity, and 75.6\% overall. All preferences were significantly above chance under an exact two-sided binomial test.

\subsection{Exploration Study}
\label{subsec:eval_user}

We also conducted a study with six participants to investigate how the design space is used. Appendix~\ref{appendix:exploration_study} gives the full study protocol and findings; we provide a summary here.

Each participant drew two sketches of their own choosing and explored one with our system and one with a VLM-only baseline, in counterbalanced order (Figure~\ref{fig:User_2}). In both conditions, the design space was generated from the participant's sketch. In the baseline, the sketch produced 16 whole-shape generations; participants chose five favorites, edited each with a text instruction, and then explored a design space formed by recombining their segmented components. Because these shapes are generated and segmented independently, their components are not guaranteed to fit together when recombined. Our system instead generates a per-component design space directly from the sketch for selection and composition.

We then conducted a semi-structured interview to understand participants' workflow preferences, how component compatibility affected exploration, and how exploration developed their initial sketches. All participants preferred selecting components to selecting whole shapes: they often liked components from several designs rather than any single generated shape. They also found selection easier than describing desired components in text, since they did not always have the words for the component they wanted. Component compatibility also mattered: floating and overlapping components in the baseline drew immediate negative reactions, and participants said these failures could make it impossible to explore a combination further.

Participants reported that exploring our design spaces often developed their rough sketches and suggested new ideas, although whole-shape variants could sometimes suggest more divergent alternatives by changing the entire design at once. Finally, participants wanted more control over segmentation granularity. Some wanted finer components while others found excessive segmentation distracting, but all requested hierarchical segmentation that would let them open a component and explore it at finer granularity. Figure~\ref{fig:User_feedback} shows designs participants produced with our system.

%% file: sections/8_Conclusion.tex
\section{Discussion and Limitations}
\label{sec:limitations}

We discuss three aspects of our current formulation that shape its scope: component decomposition, geometry representation, and the relations between components.

\paragraph*{Part decomposition.}
We currently segment each object into a single flat set of components. This fixes the granularity of exploration: a designer can vary a component, but cannot open it and explore its subcomponents independently. Participants in our exploratory study explicitly asked for this capability. Future work could support hierarchical decomposition, allowing designers to move between object-, component-, and subcomponent-level exploration.

\paragraph*{Geometry representation.}
Our components are represented as CadQuery programs, which favor prismatic operations such as profiles, extrusions, fillets, and booleans. This representation works less well for objects whose identity lies in freeform surfaces, such as car bodies; for these shapes, our variants primarily change the massing rather than the surface design. Future work could extend our representation with NURBS or other freeform operations while retaining part-level interchangeability.

\paragraph*{Constraints and interfaces.}
Our formulation supports four design relations---attachment, alignment, proportion, and mirroring---and structural constraints that keep components disjoint and make attached pairs meet over a single flat interface. This excludes curved or disconnected interfaces, as well as design requirements such as clearance, function, manufacturability, and more general symmetry. Future work could expand both the relation vocabulary and the interface representation to support a broader range of assemblies and design requirements.

\section{Conclusion}
\label{sec:conclusion}

We presented Composable CAD, a system for generating a part-based design space from a concept sketch. Generating components independently gives each component its own detail and variants, but those variants are only useful as a design space if they can be freely recombined. We make this possible by recovering the sketch's attachment, alignment, and proportion relations to guide component generation, and by deterministically rewriting the resulting programs so their components meet at shared interfaces. The resulting components are interchangeable across samples, allowing designers to explore, edit, and physically assemble different combinations without regenerating the rest of the design. We hope composable part generation opens new ways to use generative CAD for design exploration.

%% file: sections/fig_only.tex
\begin{figure*}[t!]
\includegraphics[width=\linewidth, height=6.05cm]{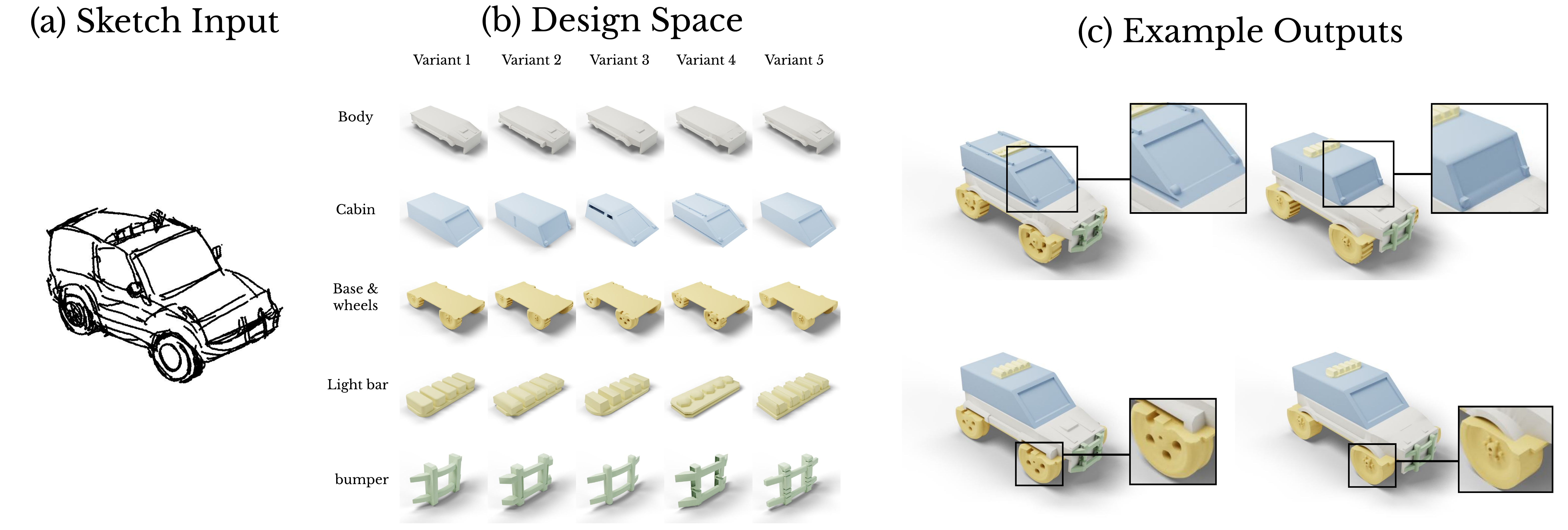}
\caption{Design space for a sketch of an off-road vehicle. From the input sketch (a) we segment the object into components and generate five samples for each, one row per component (b). The designer composes a shape by picking one sample per row, giving $5^5 = 3125$ possible vehicles; four are shown (c), with insets on the components that differ between them. Every combination assembles into a single valid solid, so the samples can be mixed freely. The remaining result figures follow this format.}
\label{fig:result_car_1}
\end{figure*}

\begin{figure*}[t!]
\includegraphics[width=\linewidth, height=6.7cm]{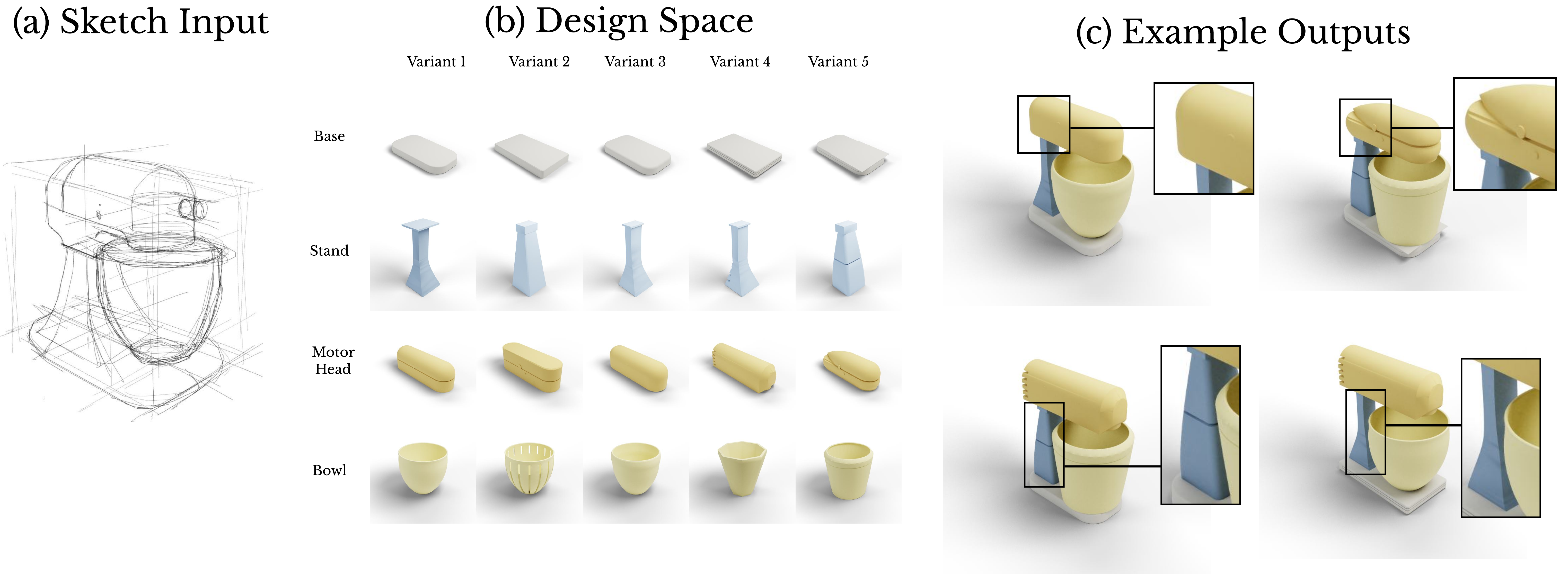}
\caption{Design space for a stand mixer. The sketch is from OpenSketch~\cite{gryaditskaya2019opensketch}, a dataset of professional product design sketches, so it carries dense construction lines and a great deal of detail. Insets highlight the components that differ between the shown outputs.}
\label{fig:result_mixer}
\end{figure*}

\begin{figure*}[t!]
\includegraphics[width=\linewidth, height=5.9cm]{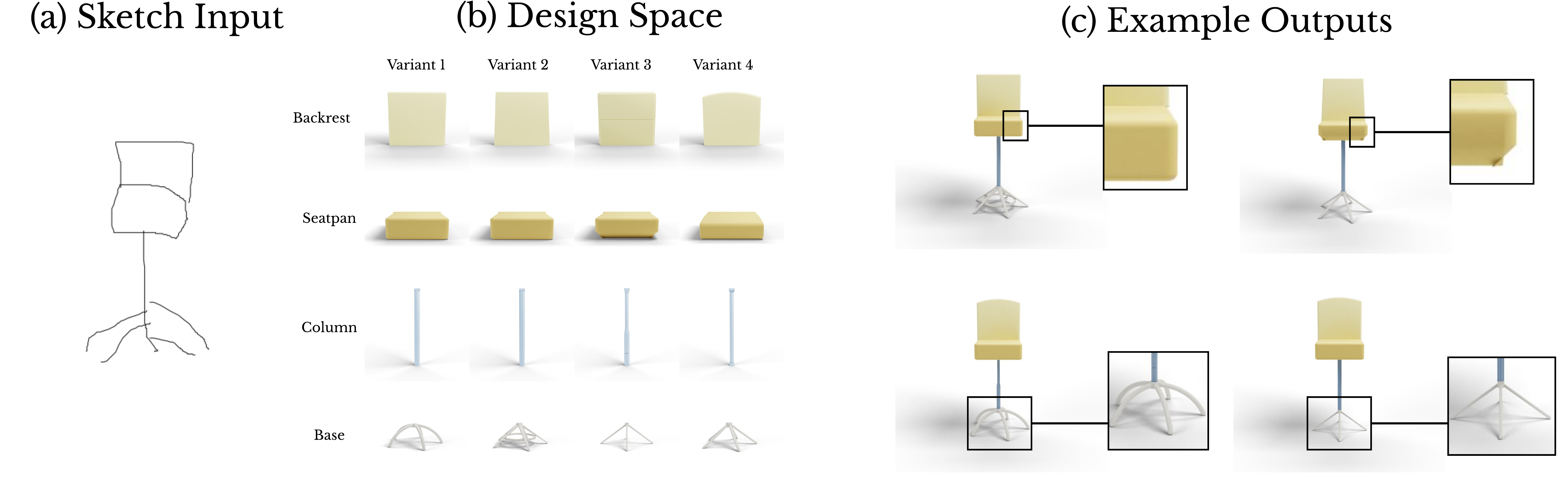}
\caption{Design space for a task chair, drawn with only a few loose strokes. Because the sketch specifies so little, the samples depart from it further and the design space is correspondingly broader. Insets highlight the components that differ between the shown outputs.}
\label{fig:result_seat}
\end{figure*}

\begin{figure*}[t!]
\includegraphics[width=\linewidth, height=6.6cm]{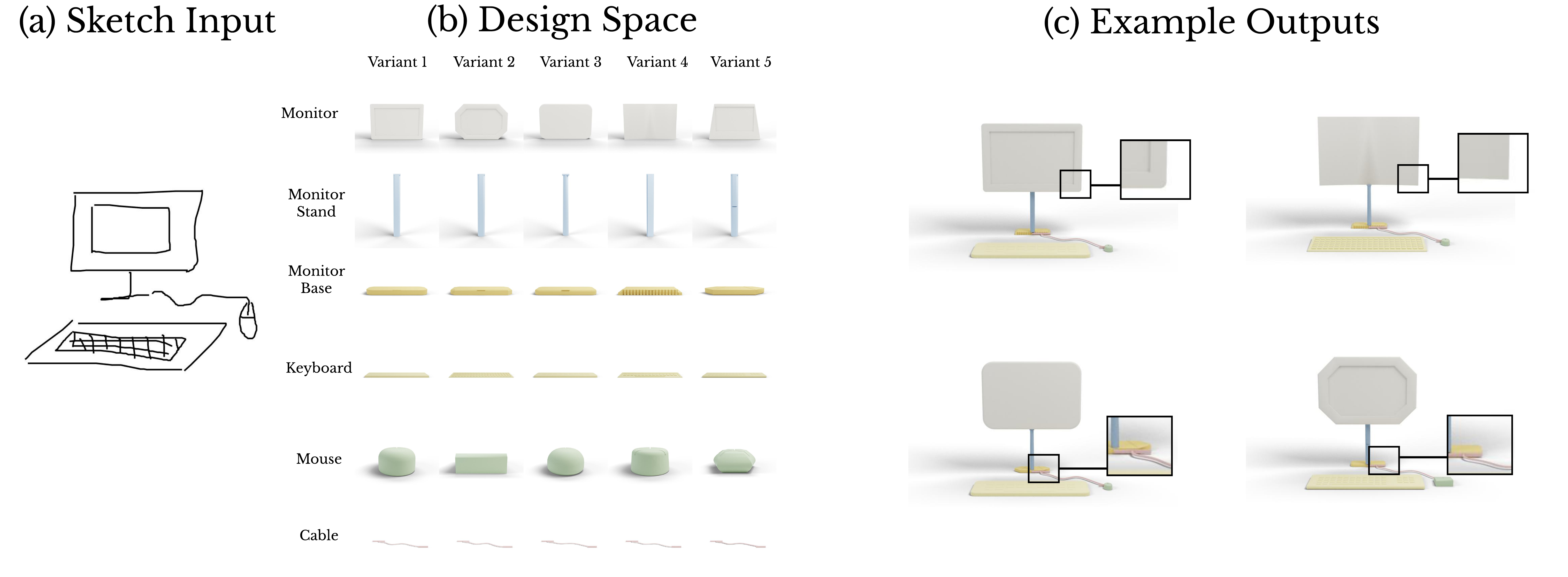}
\caption{Design space for a desktop computer, segmented into six components. Insets highlight the components that differ between the shown outputs.}
\label{fig:result_computer}
\end{figure*}

\begin{figure*}[t!]
\includegraphics[width=\linewidth, height=5.7cm]{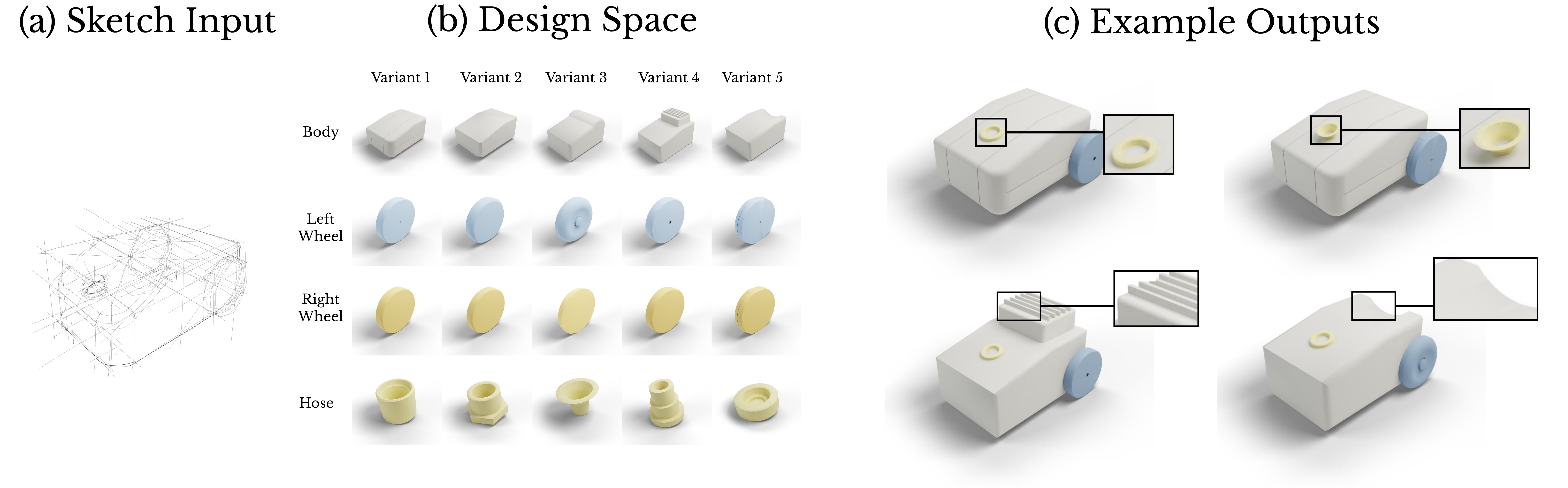}
\caption{Design space for a wheeled cart. The sketch is from OpenSketch~\cite{gryaditskaya2019opensketch}, drawn with light construction lines. Insets highlight the components that differ between the shown outputs.}
\label{fig:result_vacuum}
\end{figure*}

\begin{figure*}[t!]
\includegraphics[width=\linewidth, height=5.45cm]{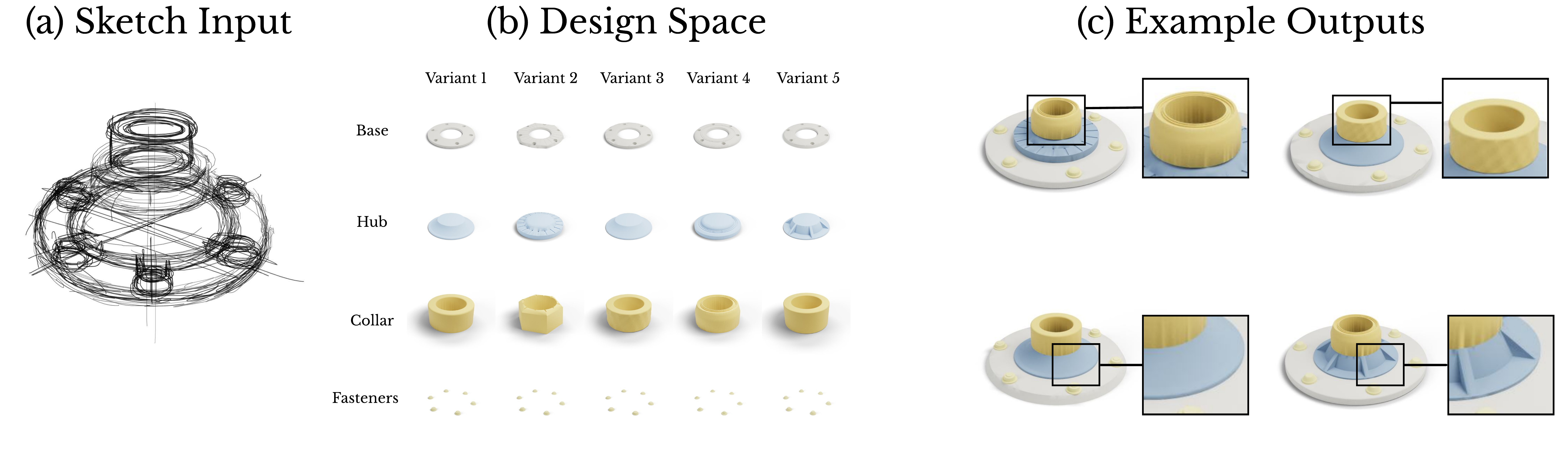}
\caption{Design space for a mechanical flange. The sketch is from OpenSketch~\cite{gryaditskaya2019opensketch}, drawn with heavy over tracing that leaves the outlines ambiguous. Insets highlight the components that differ between the shown outputs.}
\label{fig:result_flan}
\end{figure*}

\begin{figure*}[t!]
\includegraphics[width=\linewidth, height=3.4cm]{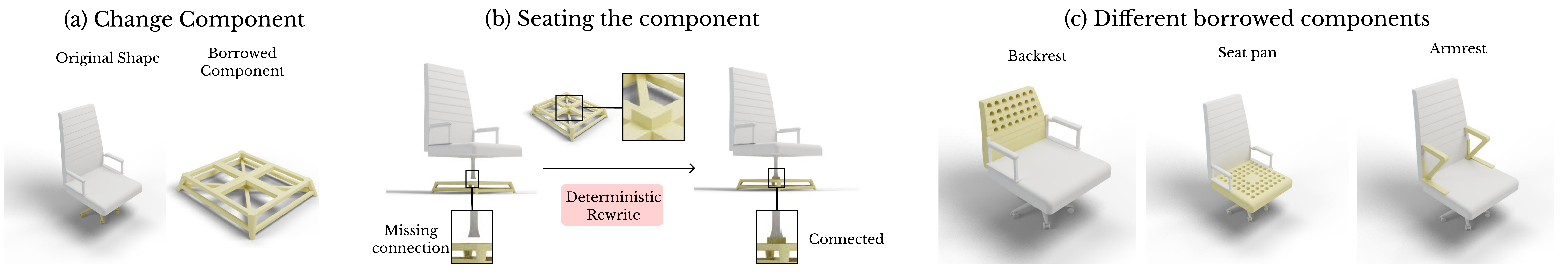}
\caption{Editing a component in 3D. Instead of drawing a part, the designer borrows a finished one from another design (a), here a base at a different size and proportion than the chair it will join. Placed directly it does not meet the chair (b, left inset); we assign it the connecting surfaces of the part it replaces and run the deterministic rewrite, which rescales and seats it flush (b, right inset). Other borrowed parts, a backrest, a seat pan, and a pair of armrests, seat into the chair the same way (c).}
\label{fig:3D_Edit}
\end{figure*}

\begin{figure*}[t!]
\includegraphics[width=\linewidth, height=3.75cm]{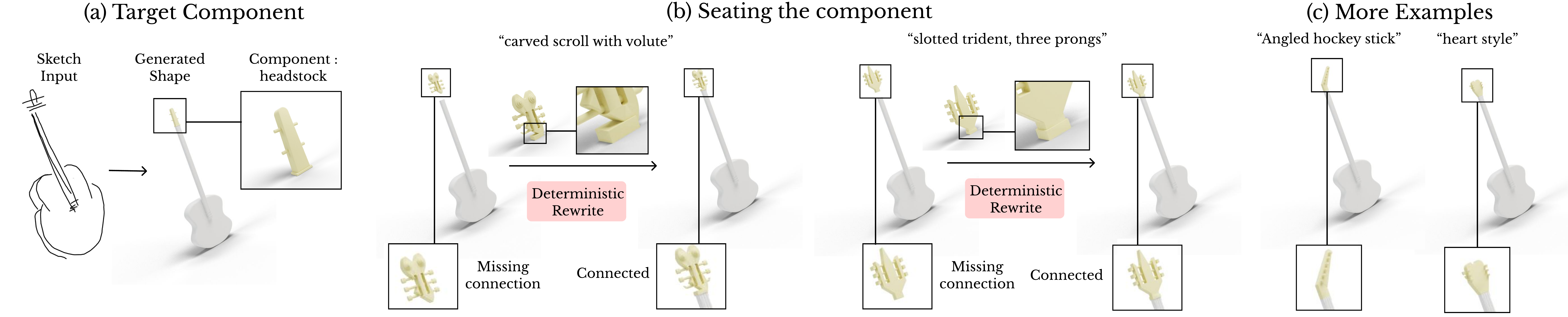}
\caption{Editing a component from a text prompt. Instead of drawing or borrowing a part, the designer describes it, and the model generates a headstock from the prompt (a). Placed directly it misses the neck (b, left inset); assigning it the connecting surfaces of the part it replaces and running the deterministic rewrite seats it flush (b, right inset). Different prompts yield different headstocks, each seating into the same fixed guitar (c).}
\label{fig:text_edit}
\end{figure*}

\begin{figure*}[t!]
\centering
\includegraphics[width=\linewidth, height=13.5cm, keepaspectratio]{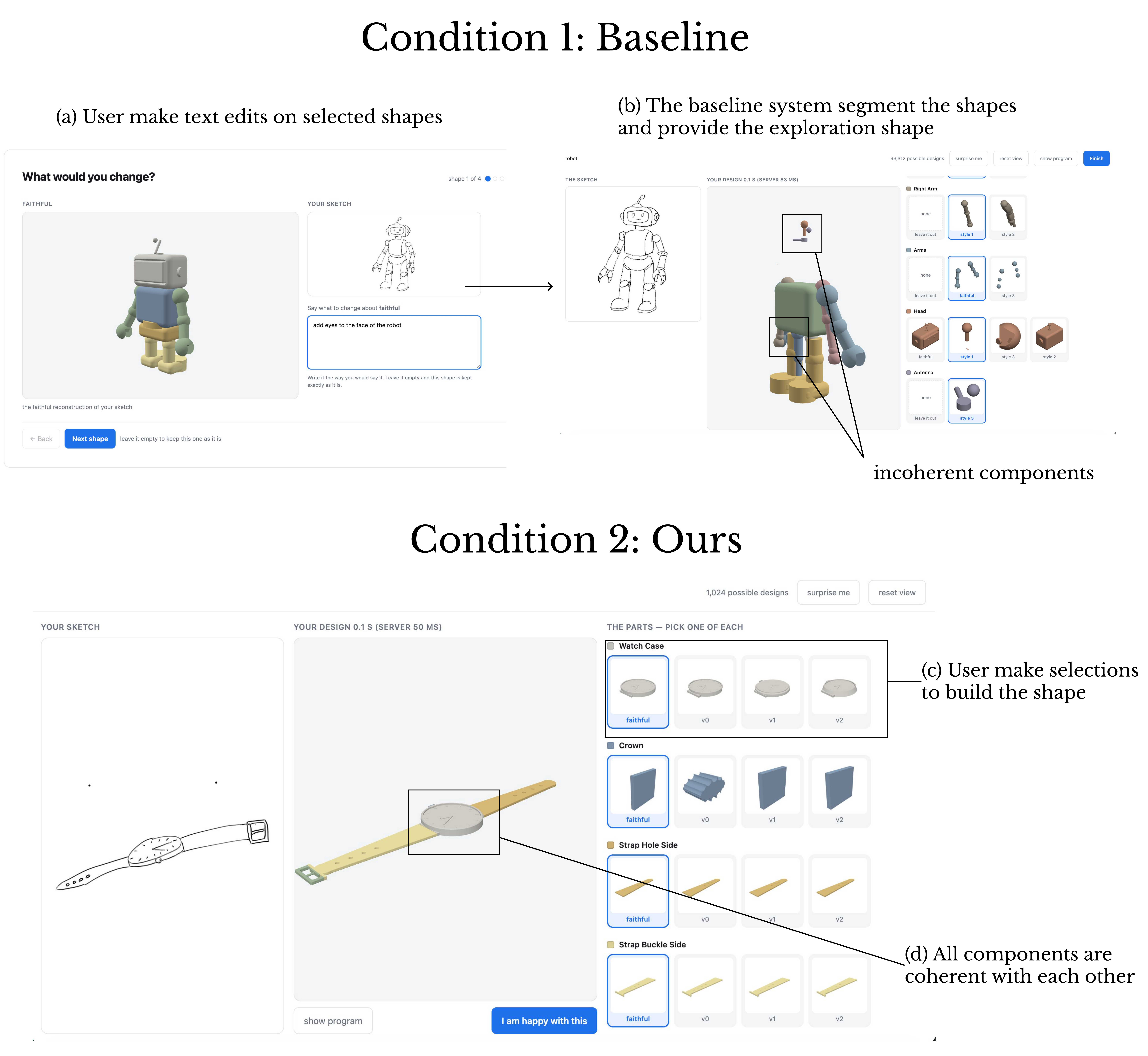}
\caption{The two conditions in the exploration study. In the baseline, the participant issues a single text instruction on each shape they picked from the gallery (a), and the system then segments those shapes and presents them as a design space to combine from (b). Because each shape was generated and segmented on its own, the parts do not meet: the insets show an arm and a head floating clear of the body. Our condition presents a per-component design space directly (c), and any selection composes into one solid, so the parts meet flush wherever they join (d).}
\label{fig:User_2}
\end{figure*}

\begin{figure*}[t!]
\includegraphics[width=\linewidth, height=11.5cm]{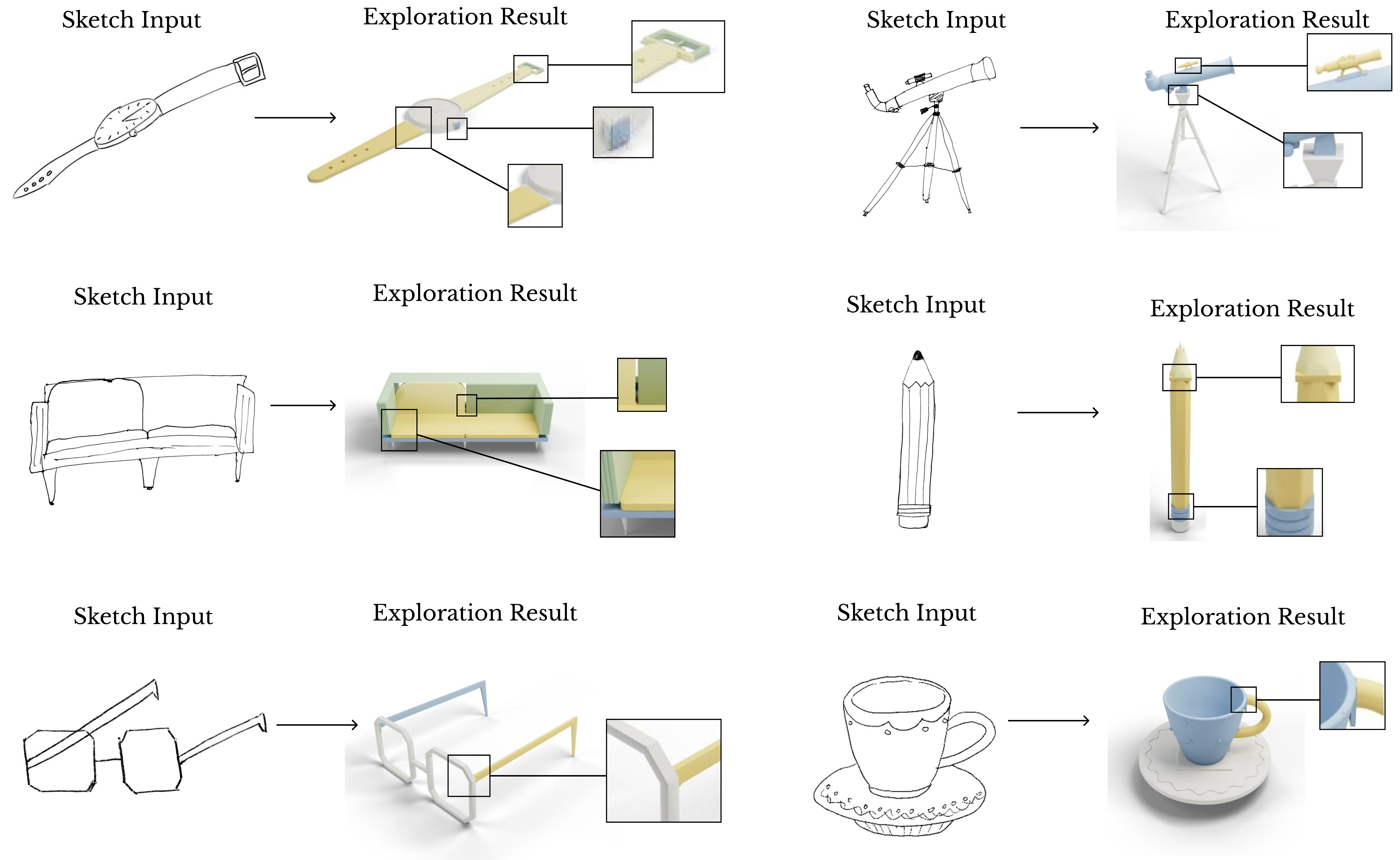}
\caption{Shapes participants produced with our system during the study. Each participant drew a sketch on a tablet (left) and composed a shape from the resulting design space (right). The insets show where components meet: the strap and case of the watch, the barrel and mount of the telescope, the arms and frame of the glasses, the cup and its handle. Every joint is flush, with no gap or overlap, although each component was generated on its own and the participant was free to combine any sample with any other.}
\label{fig:User_feedback}
\end{figure*}

%% file: sections/9_Appendix.tex
\section{Formative Study}
\label{appendix:formative_study}

\begin{table}[t]
\centering
\begin{tabular}{lll}
\hline
Role & Specialization & Location \\
\hline
Professional & Design studio & New York \\
Professional & Freelance design & Shanghai \\
Student &  Industrial Design/ME & Colombia \\
\hline
\end{tabular}
\caption{Formative study participants. All were industrial designers; institution names are anonymized.}
\label{tab:formative_participants}
\end{table}
To ground our system design in real practice, we conducted semi-structured interviews with three industrial designers spanning a range of career stages and specializations (Table~\ref{tab:formative_participants}). We report two findings.
\paragraph{Design has two phases.}
Participants described their work as two phases: an early phase of brainstorming and loose sketching, where they explore many ideas quickly, and a later phase of building and adjustment, where they realize the shape in 3D and vary its parameters to produce different designs. Our system targets the second phase, where designers want to build the shape from the input sketch quickly while also exploring variants of it.
\paragraph{Exploration is the bottleneck.}
Participants told us that producing variants is time-consuming. Adjusting parameters alone is straightforward, but the variants worth exploring usually require changing the program itself, adding or reworking detail, which is far more effort. This is the work our system aims to take on for them.
\paragraph{Feature Requests}
We list the core capabilities participants asked for, which motivated our system design and are what our system is built to provide.
\begin{enumerate}
\item Quick building from a sketch. Given the input sketch, the designer can obtain the 3D shape quickly.
\item Imagine and fill in. Where the sketch is loose or silent, the system plausibly completes what it leaves unspecified.
\item Part-based exploration in 3D. Designers explore in 3D without committing to a single idea, so the system offers alternatives per part that recombine freely rather than one whole-shape variant.
\item Multi-part compatibility. Independently explored parts stay mutually compatible, so a designer, or a collaborator owning one part, can revise a part and still assemble a valid whole.
\end{enumerate}

\section{Sketch Segmentation}
\label{appendix:sketch_segmentation}
Sketch segmentation has been widely studied~\cite{yang2021sketchgnn,qi2022onesketch,tang2025inklayer}. We segment the input sketch in two API calls: one to identify semantic parts and one to generate a color-coded segmentation mask. We first give the sketch to Gemini~3.1~Pro~\cite{gemini31pro2026} and prompt it to list the object's semantic parts; for the chair, these are the base, column, seat pan, backrest, and the two armrests. The part-list prompt is:
\begin{quote}
\itshape
You are given a single freehand concept sketch of one object. List the object's semantic parts: the parts a designer would model and vary separately. Output one part per line, as a short lowercase noun label, and nothing else. Merge duplicate instances only if they are identical; otherwise label them separately (e.g. ``left armrest'', ``right armrest'').
\end{quote}
We then provide the sketch, part list, and color legend to Gemini~3~Pro~Image~Preview (\texttt{gemini-3-pro-image-preview})~\cite{geminiImageGeneration2026}, requesting a color-coded mask with the same dimensions as the input. We use the image returned by the API as the segmentation mask and recover each part's region by extracting the pixels of its assigned color. The mask prompt is:
\begin{quote}
\itshape
You are given the same sketch and a list of its parts: \{part\_list\}. Return a single image with the exact same width and height as the input. Paint the region of each part in a solid, flat, distinct color, using this assignment: \{color\_legend\}. Fill each part's full area, keep boundaries tight to the drawn strokes, and paint the background pure black (0,0,0). Do not draw outlines, labels, gradients, shading, or anti-aliased edges: use only the exact assigned colors and black.
\end{quote}

\section{Geometric Prior}
\label{appendix:geo_relations}

\begin{figure}[t!]
  \includegraphics[width=\linewidth]{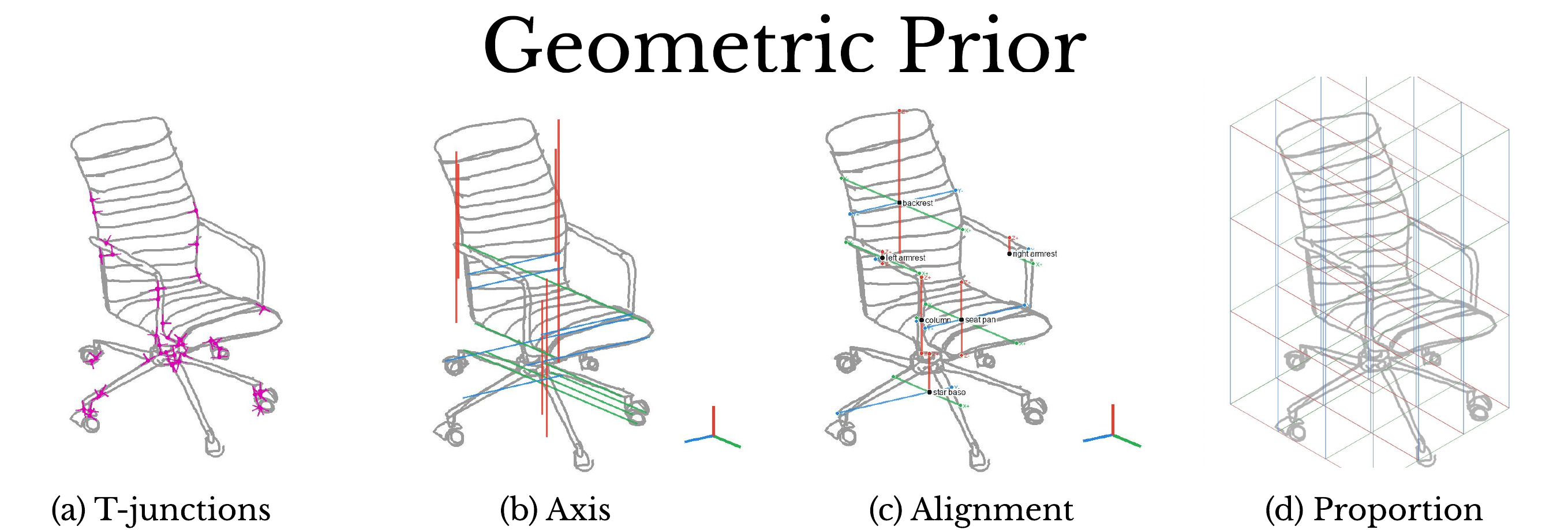}
\caption{Geometric prior extraction, illustrated on a chair sketch. (a) T-junctions detected among the strokes (magenta), marking where parts cross. (b) The recovered coordinate axes: world $z$ from the page vertical (red), and the two horizontal axes voted from the junction-incident stroke directions (blue, green). (c) Alignment relations, where part faces and centers that coincide along an axis are linked, with the participating parts labeled. (d) Per-part proportions, read as the $x{:}y{:}z$ extents of each part's box in the recovered frame.}
\label{fig:geo_prior}
\end{figure}

We recover the design constraints from the vectorized strokes and the part masks in four deterministic passes. Junctions tell us where strokes cross and which parts touch; from the junctions we recover the object's coordinate axes; and against those axes we compute the alignment and proportion relations, since both are defined per axis. We first crop the sketch to its drawn content and resize it into a $256 \times 256$ box, with a margin of 20 pixels on each side, so every sketch is processed at the same resolution and all thresholds below are stated in pixels of that box.

\paragraph{Junctions.}
A junction is where one stroke meets another, and it is the primary 2D cue for which parts contact (Figure~\ref{fig:geo_prior}a). For an endpoint of stroke $A$ meeting stroke $B$, we register a T-junction when the strokes differ in direction by at least $25^\circ$ (so a collinear break is not counted), the endpoint projects to the interior of $B$ at parameter $t \in [0.15, 0.85]$ (a mid-stroke hit, not $B$'s endpoint), and the perpendicular distance from the endpoint to $B$ is at most 8px. Near-duplicate junctions are merged within small pixel clusters. We then attribute each junction to parts with a KD-tree over the masks, assigning it to its two nearest parts when both lie within 13px. Votes accumulate over unordered part pairs, and every voted pair emits an attachment relation, promoted to a resting relation when the vertical offset dominates ($dy > 0.6dx$).

\paragraph{Axes.}
The relations we extract are all stated per world axis, so we next recover the object's coordinate frame (Figure~\ref{fig:geo_prior}b). The vertical axis is fixed by convention: designers draw the up-direction vertical on the page, so we take world $z$ to project onto the image's vertical direction. The two horizontal axes are recovered by voting with the junctions, since at each junction the two crossing strokes are corner edges rather than arbitrary marks and so are strong evidence for the axis directions. We collect the direction of every junction-incident stroke, drop those within $15^\circ$ of vertical, and cluster the rest by angle, each junction contributing a vote to its cluster. The two highest-voted clusters give the horizontal axis directions, which we orthogonalize against $z$ to form the frame. A part's extent along an axis is then read by projecting its mask onto that axis direction, which is what the following two passes use.

\paragraph{Alignment.}
With the axes fixed, we find which parts line up along each (Figure~\ref{fig:geo_prior}c). Alignment is stated between anchors: on a given axis, a part contributes three of them, its minimum, its center, and its maximum, so that a designer's flush edge, shared centerline, or edge-to-center placement are all expressed in the same way. We project every part's mask onto the axis, collect the anchors of all parts, and cluster their values by complete linkage: a new cluster starts whenever a value exceeds the current cluster's minimum by more than 7px, so each cluster spans at most 7px and no chaining occurs. A cluster holding anchors from at least two distinct parts is kept, and each pair of parts within it emits an alignment relation on that axis, recording which anchor each side contributed and the residual offset between them.

\paragraph{Proportion.}
For each part we compute an $x{:}y{:}z$ shape hint from its extent along the three axes (Figure~\ref{fig:geo_prior}d), read in the recovered frame rather than from the silhouette, since a single isometric view cannot separate the three axes. We divide the three extents by the longest, so the longest axis is 1.0, and snap each ratio to a 0.1 grid floored at 0.1.

\section{Design Constraints as Functions}
\label{appendix:constraints_as_functions}

\begin{figure}[t!]
  \includegraphics[width=\linewidth]{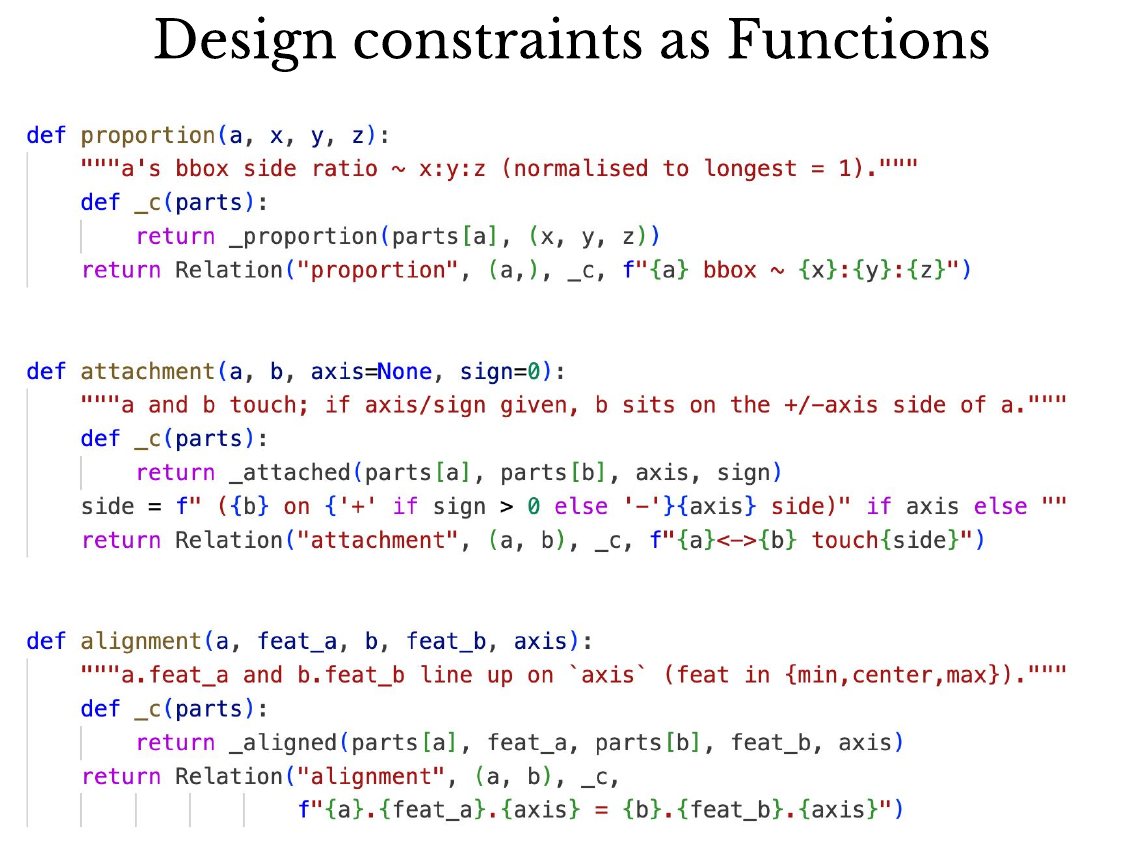}
\caption{Design constraints compiled as functions. Each relation extracted from the sketch becomes a \texttt{Relation} carrying a predicate that is run against the executed program, returning whether the relation holds and by how much.}
\label{fig:design_constraints_func}
\end{figure}

Each design constraint from the sketch is compiled into a small function over the generated program, so that checking a relation is just executing the programs. Figure~\ref{fig:design_constraints_func} shows the three stroke-derived design constraints: \texttt{proportion}, \texttt{attachment}, and \texttt{alignment}; mirroring comes from the segmentation and is checked the same way. To evaluate a program, we execute it to obtain the parts, then call each relation's relation function on them. The result is a per-relation pass or fail with its residual, which is the report returned to the VLM (Section~\ref{sec:overview}) as natural language feedback.

\section{Proxy Generation}
\label{appendix:proxy_generation}
We generate the initial proxy with Gemini 3.1 Pro (\texttt{gemini-3.1-pro})~\cite{gemini31pro2026}, giving it the input sketch and the prompt:
\begin{quote}
\itshape
You are given a concept sketch of a single object and the list of its parts: \{part\_list\}. Write a CadQuery program that reconstructs the object as a proxy. Represent each part as one simple primitive volume, and define exactly one function per part, each named after its part and returning that part's solid (for example, \texttt{def backrest() -> cq.Workplane}). Place every part in a shared coordinate frame, sized and positioned so that the proxies together match the object in the sketch: its proportions, the way the parts are aligned, and which parts meet. Do not add detail or features; one primitive per part is enough. Render your program and compare the result against the sketch before finalizing. Return only the CadQuery program, with no explanation.
\end{quote}
We run the model with medium thinking, and let it render its program to 2D from four isometric views and compare the renders against the input sketch, revising until the two agree. We cap this initial pass at a three minute budget and stop when it is reached. The result is the initial proxy, which we then check against the design constraints and return as feedback (Section~\ref{sec:overview}).

\section{Sampling}
\label{appendix:sampling}

We generate several samples per part with Gemini 3.1 Pro (\texttt{gemini-3.1-pro})~\cite{gemini31pro2026}, giving it only that part's proxy program, the CadQuery code that produces its proxy volume, and the prompt:
\begin{quote}
\itshape
Below is a CadQuery program that produces a simple proxy volume for the \{part\} of a \{object\}. Edit it into a detailed, realistic version of the same part. Keep the part's overall placement and extent close to the proxy, and add the geometry, features, and detail the proxy leaves out. Refine and extend the existing operations rather than rewriting the program from scratch. Return only the edited CadQuery program, with no explanation.
\end{quote}
Because the input is CadQuery code rather than a description, the model works by editing the existing program, refining and extending the operations that produce the proxy, not writing a new one. We draw five samples per part from independent calls and control their diversity through the sampling temperature, using 0.1, 0.2, 0.3, 0.4, and 0.5, so the lowest stays close to a faithful elaboration of the proxy and the highest departs furthest from it. All parts and all samples are generated in parallel.

\section{Cut or Union: Why the Choice Depends on the Surface}
\label{appendix:suface_analysis_counter_case}

\begin{figure}[t!]
  \includegraphics[width=\linewidth]{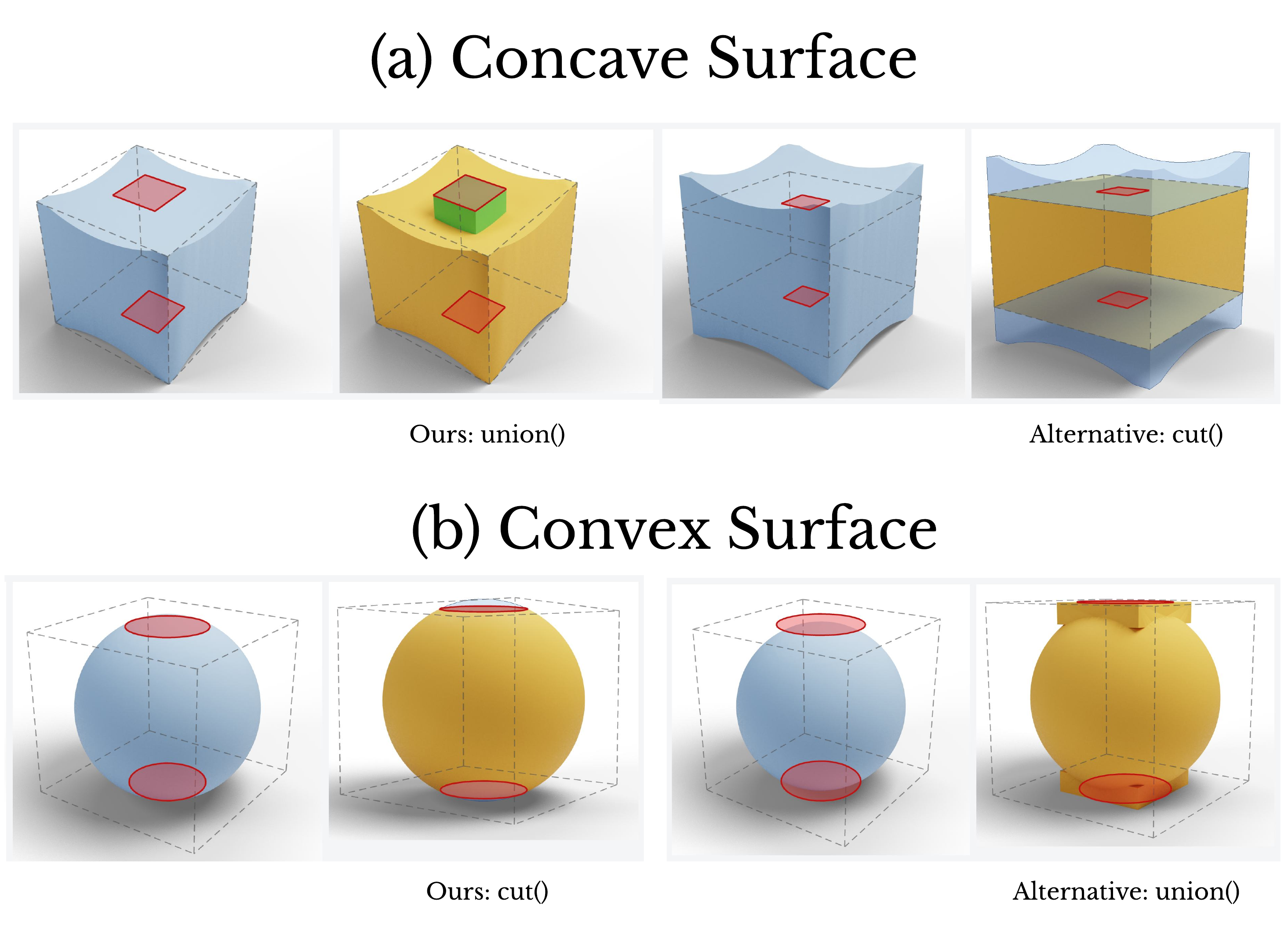}
\caption{The two surface types and why the choice between cut and union depends on them. Each row shows a part that has already been translated and scaled to reach its connecting surfaces (red); only the last operation differs. (a) A concave surface falls short of the plane. Our union fills the gap (left), while the alternative, a cut, must slice the part down to where its surface already reaches, flattening the whole face and removing everything above it (right). (b) A convex surface protrudes past the plane. Our cut trims the thin excess (left), while the alternative, a union, wraps material around the bulge up to the plane, replacing the curved face with a flat one (right). In both cases the alternative reduces the surface to a bare plane and discards its features.}
\label{fig:alt_rewrite}
\end{figure}

To form a flat face at a connecting surface, the rewrite either cuts material away or adds it. We choose the operation based on how the part's surface sits relative to the connecting surface. Taking the wrong operation destroys the geometry the sampling produced; Figure~\ref{fig:alt_rewrite} shows the counterexamples.

For a concave surface, one that falls short of the connecting surface, we union: this fills only the shallow gap and leaves the surface intact. A cut cannot trim there, since the material is already below it; to expose a flat face it must slice the part down to the lowest point the surface reaches, which flattens the whole footprint into a bare plane and removes every feature that sat above it. For a convex surface, one that protrudes past the connecting surface, the situation is reversed and we cut, removing only the thin part that overshoots and leaving the rest of the surface as it was. A union instead wraps added material around the protruding geometry, so the curved face is buried and the interface becomes a flat block, again erasing the feature. In both cases the alternative reduces a shaped surface to a featureless plane, so we take the operation that keeps the surface and changes the part as little as possible.

We classify the surface locally, only based on the geometry at the connecting surface. Over the region where the two parts meet, we sample the part's surface and measure how far it reaches past the connecting surface, calling it convex when the largest such distance exceeds a small tolerance and concave otherwise.

\section{Proof of the Composability Proposition}
\label{appendix:proof}

We make the setting of the proposition explicit. Every proxy $B_c$ is an axis-aligned box, and the proxies have pairwise disjoint interiors. Each connecting surface $s_i$ is a planar region of positive area on a face of $B_c$ and is the entire intersection $B_c\cap B_{c'}$ with its attached neighbor. The two proxies lie on opposite sides of its plane $\pi_i$. The attachment graph is connected, and the proxies and interfaces are fixed across samples.

Let $\widetilde V_c$ be the solid after Steps~1 and~2. We assume that its material inside $B_c$ forms a nonempty connected solid. At every interface handled by a convex cut, this in-box material already contains the entire $s_i$. We also assume that every part of $\widetilde V_c$ outside $B_c$ lies in a region removed by one of the convex cuts, and that each concave extrusion produces a solid inside $B_c$ that meets the part when added. These assumptions restrict the samples to which the proposition applies.

Step~3 retains the two cases of Algorithm~\ref{alg:rewrite}. At a convex surface, the cut removes material on the outward side of $\pi_i$. The material already covering $s_i$ lies inside $B_c$ and is retained, so the cut exposes a flat face containing $s_i$. At a concave surface, the added extrusion has $s_i$ as a face in $\pi_i$, so its union with the part also covers $s_i$. Later operations preserve this coverage: every cut acts outside $B_c$, and a union cannot remove an existing interface.

All material added by the unions lies inside $B_c$, and all original material outside $B_c$ is removed by the cuts. Hence the final output satisfies $V'_c\subseteq B_c$, establishing (i). The two branches establish $s_i\subseteq V'_c\cap\pi_i$ for every $i$, giving (ii). Since each $\pi_i$ supports $B_c$, containment also gives (iii): no material lies on the outward side of the interface plane.

The output is connected as well. The cuts leave the original in-box solid intact. Each added extrusion lies inside $B_c$ and meets the current part, so it meets the current in-box material. Thus every union preserves the connectedness of that material. Once all exterior material has been removed, the remaining solid is therefore connected.

The composition properties follow. Every selected solid lies inside its own proxy, so their interiors are pairwise disjoint. For an attached pair $c,c'$ sharing $s_i$,
\[
s_i
\subseteq V'_c\cap V'_{c'}
\subseteq B_c\cap B_{c'}
=s_i.
\]
Thus the pair meets exactly along $s_i$, with no gap or volumetric overlap. Every selected solid is connected, and the connected attachment graph joins them through these shared interfaces, so their union is connected.

The argument applies to any selection of samples satisfying the stated assumptions for which the rewrite completes within the material budget. The budget does not weaken the resulting geometric properties. \hfill$\square$

\section{Joints}
\label{appendix:Joints}

\begin{figure}[t!]
  \includegraphics[width=\linewidth]{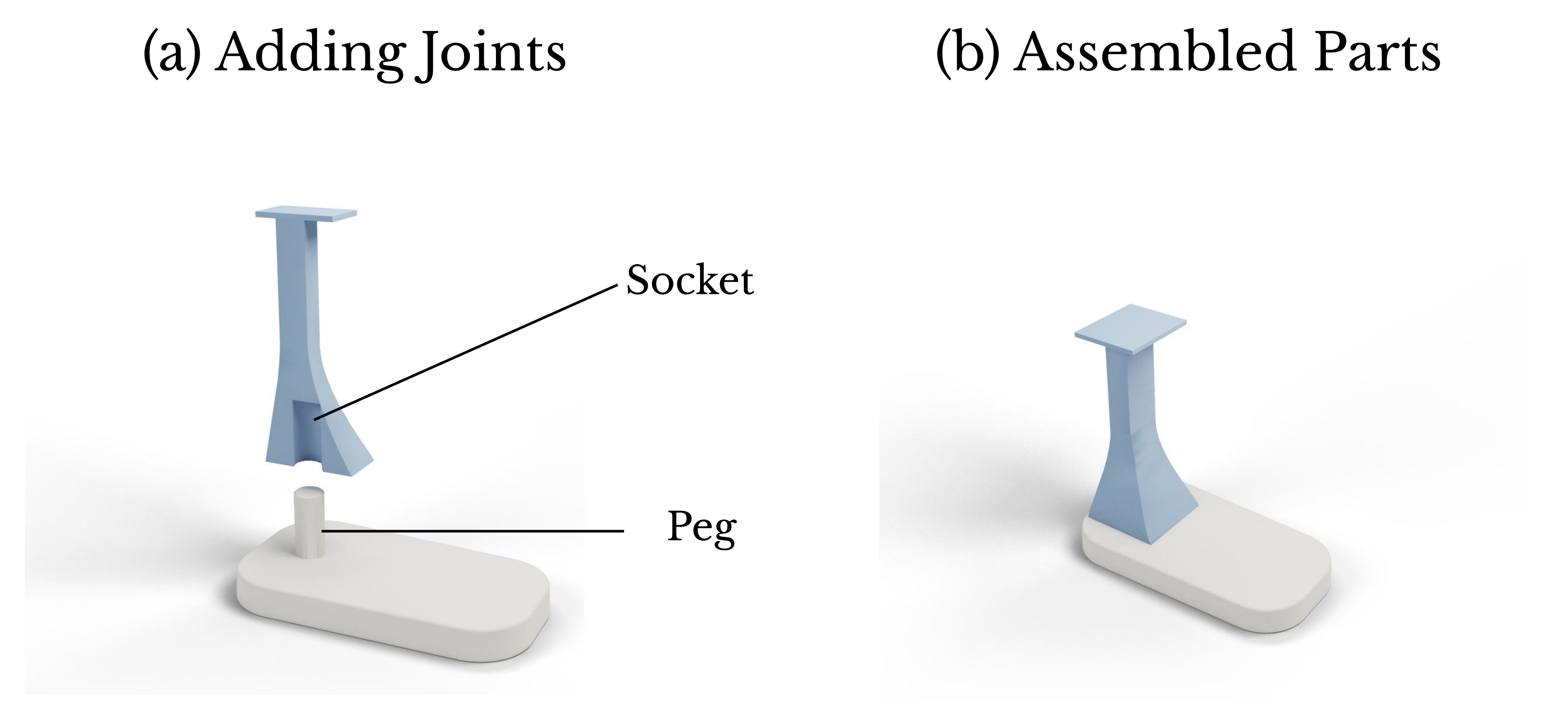}
\caption{Joints between parts. (a) At each connecting surface we add a peg on one part and a matching socket on the other, both centered on the surface and running along its normal. (b) The two parts assembled, with the joint hidden inside and the outer geometry undisturbed.}
\label{fig:joints}
\end{figure}

To hold assembled parts together we add a peg-and-socket joint at every connecting surface (Figure~\ref{fig:joints}). The joint is centered on the connecting surface and runs along its normal, so it is hidden once the parts are assembled and does not disturb the outer geometry.

We first choose which side takes the socket. Drilling along the normal is limited by how much material lies behind the surface on each side, so we take the side that admits the deeper hole and place the socket there; the matching peg is added to the other part. The socket diameter is $\tfrac{1}{2}\max(0.8\,\ell,\ 0.5\text{ cm})$, where $\ell$ is the diagonal of the connecting surface, so the joint scales with the interface but never falls below a printable size. Its depth is the smaller of $2$ cm and the depth the material behind the surface permits. The peg is generated as the same cylinder on the opposite part, so the two are congruent by construction.

\section{Setup Prompts}
\label{appendix:setup_prompts}

The prompts for the two VLM baselines of Section~\ref{subsec:evaluation_setup}. The prompt for $S_1$, which produces the whole object in one pass, is:
\begin{quote}\small\itshape
You are given a concept sketch of a single object. Write one CadQuery program that reconstructs the entire object. First decide the object's semantic parts, the parts a designer would model separately. Then define exactly one function per part, each named after its part and returning that part's solid (for example, \texttt{def backrest() -> cq.Workplane}). Place every part in a shared coordinate frame so that together they form the object shown in the sketch, with adjacent parts meeting where they should and the proportions matching the drawing. Return only the CadQuery program, with no explanation.
\end{quote}

$S_2$ uses two prompts. The first asks for the proxy:
\begin{quote}\small\itshape
You are given a concept sketch of a single object. Write a CadQuery program that reconstructs the object as a proxy. First decide the object's semantic parts, the parts a designer would model separately. Represent each as one simple primitive volume, and define exactly one function per part, each named after its part and returning that part's solid (for example, \texttt{def backrest() -> cq.Workplane}). Place every part in a shared coordinate frame, sized and positioned so that the proxies together match the object in the sketch: its proportions, the way the parts are aligned, and which parts meet. Do not add detail or features; one primitive per part is enough. Return only the CadQuery program, with no explanation.
\end{quote}
The second is issued once per part, with the full proxy program as context:
\begin{quote}\small\itshape
Below is a CadQuery program that produces a simple proxy of a \{object\}, with one function per part. Rewrite the function for the \{part\} into a detailed, realistic version of that part, and leave every other function unchanged. Keep the part's overall placement and extent close to the proxy, and add the geometry, features, and detail the proxy leaves out. Make sure the part still meets its neighbors where the proxy has them meet. Return only the edited CadQuery program, with no explanation.
\end{quote}

\section{VLM-Based Repair Prompt}
\label{appendix:vlm_repair}

We provide the complete CadQuery program for each selected shape and use the following prompt. The execution harness enforces a three-minute wall-time limit per shape.

\begin{quote}
\itshape
You are given a CadQuery program that generates a multi-part object. Execute the program and inspect the resulting geometry. Repair gaps and overlaps between parts, and ensure that parts intended to connect meet over flat contact regions. Preserve the existing part structure and geometric details, including holes, curved features, and thin elements. Keep the parts as separate functions so that they remain individually accessible. You may execute and revise the program repeatedly. Return the final executable CadQuery program.
\end{quote}

\section{Perceptual Comparison Protocol}
\label{appendix:User_protocol_1}

We ran a pairwise comparison to test whether the extra geometry our programs carry is preferred by people, since program length alone does not establish that it is worth having.

\paragraph*{Participants.}
We recruited 12 participants from within our organization by email. Participants had to be at least 18 years old and proficient in English, and were not required to have professional experience in sketching, 3D modeling, or generative AI. Each session lasted about 30 minutes and participants were compensated with a small token of appreciation.

\begin{figure*}[t!]
\includegraphics[width=\linewidth, height=7.5cm]{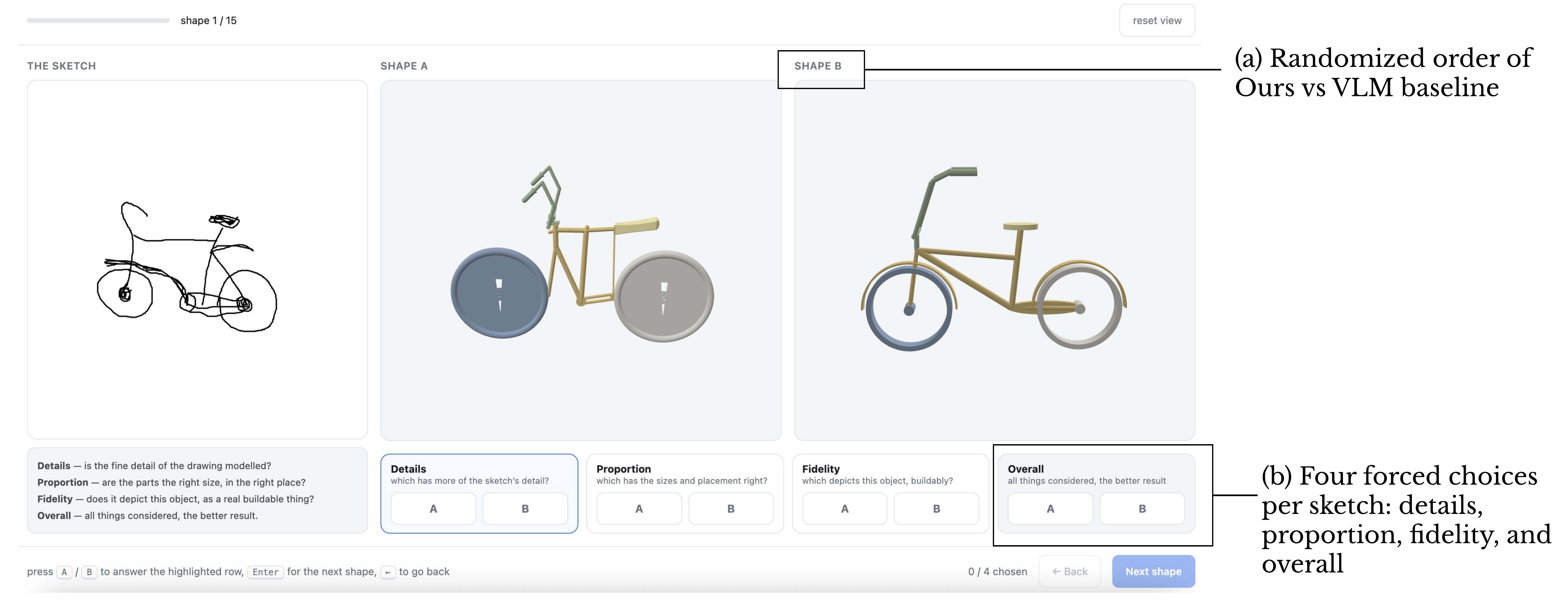}
\caption{The comparison interface. Each trial shows the input sketch on the left and two shapes generated from it, one by the single-pass baseline and one by our pipeline, in randomized left and right order and unlabeled (a). Participants can rotate either shape freely, and answer four preference questions before continuing to the next sketch (b).}
\label{fig:User_1}
\end{figure*}

\paragraph*{Stimuli.}
Each trial shows one input sketch alongside two shapes generated from it, one by the single-pass baseline ($S_1$) and one by our pipeline (Figure~\ref{fig:User_1}). The two are presented side by side as Shape A and Shape B, unlabeled, with left and right positions randomized per trial. Participants can drag either shape to rotate it and reset the camera to the starting angle at any time. Each participant sees 15 sketches drawn at random from the corpus.

\paragraph*{Task.}
Participants are told that the two shapes came from two different automatic methods, without being told which produced which, and are shown how to inspect a shape in 3D. For each sketch they answer four preference questions, with no option to abstain and no correct answer:
\begin{enumerate}
\item Which shape better preserves the details of the sketch?
\item Which shape better preserves the size and proportions of the parts in the sketch?
\item Which shape better depicts the sketched object in a buildable form?
\item Which shape do you prefer overall?
\end{enumerate}
All four must be answered before the next sketch is shown. We report, for each question, the proportion of trials on which our shape was chosen.

\section{Exploration Study}
\label{appendix:exploration_study}

We ran a user study with six participants to investigate how the design space is used. Sessions ran about an hour in person. Each participant drew two sketches on a pen tablet, on subjects of their own choosing, and explored shapes through two different systems in counterbalanced order, one sketch per condition. Design spaces were generated on the spot. Both interfaces are shown in Figure~\ref{fig:User_2}.

\textbf{The baseline condition} is a VLM-only system. A sketch produces a gallery of 16 generated shapes, from which the participant chooses their five favorites and edits them with a single text instruction before committing. We then decompose the edited shapes and present the result as a design space for the participant to combine from. Because these shapes are generated separately and segmented separately, their parts are not guaranteed to fit together.

\textbf{Ours} produces a per-part design space from the sketch, and the participant selects parts to combine, as introduced in Section~\ref{subsec:design_space}.

Each condition allowed up to ten minutes of exploration, excluding generation wait time, and ended with the participant committing to a final shape. Sessions closed with a semi-structured interview about which workflow they preferred, what they liked and disliked about each, how clearly they had the shape in mind when they drew the sketch, and whether exploring in 3D changed what they wanted. The study asks three questions: whether participants want part-based selection and exploration, how much part connectivity matters to the experience, and how far the final shape departs from the initial sketch. We report what we found on each below, and show the shapes participants produced with our system in Figure~\ref{fig:User_feedback}.

\paragraph*{Part selection was the feature participants valued.}
All participants preferred choosing parts to choosing whole shapes, and the reason they gave was consistent: no single generated shape was their favorite, but pieces of several were. As one participant put it, ``with so many options I cannot choose one favorite shape. I like a part of all of them.'' Participants also found selecting a part easier than describing one in text, since they often lacked the words for the part they wanted, and picking it from the design space asked for none.

\paragraph*{Coherent parts matter.}
In the baseline condition, where every shape is generated on its own, participants met floating and overlapping parts. These drew immediate negative reactions whenever they appeared, and participants said such parts made it impossible to explore any further. Our rewrite is what removes them, and the reaction confirms that the failure matters perceptually and not only geometrically. Participants consistently praised our condition for producing clean shapes.

\paragraph*{Exploration goes beyond the sketch.}
Participants described their own sketches as rough and unfinished, and said that exploring with our system often gave them new ideas, welcoming the additions the final shape made to their design. For some objects, however, they felt the baseline's whole-shape variants suggested more ideas than our per-part samples, since varying every part at once departs further from the sketch.

\paragraph*{Segmentation granularity needs to be controllable.}
Participants said the segmentation did not always divide the object the way they wanted. Three asked for finer parts, while one found that too many parts added noise and got in the way of a clean exploration. All of them asked for a hierarchical segmentation instead, so that a part could be opened up and explored in more detail when they chose to.

\section{Example Programs}
\label{appendix:example_programs}
\begin{figure*}[t!]
\includegraphics[width=\linewidth, height=10cm]{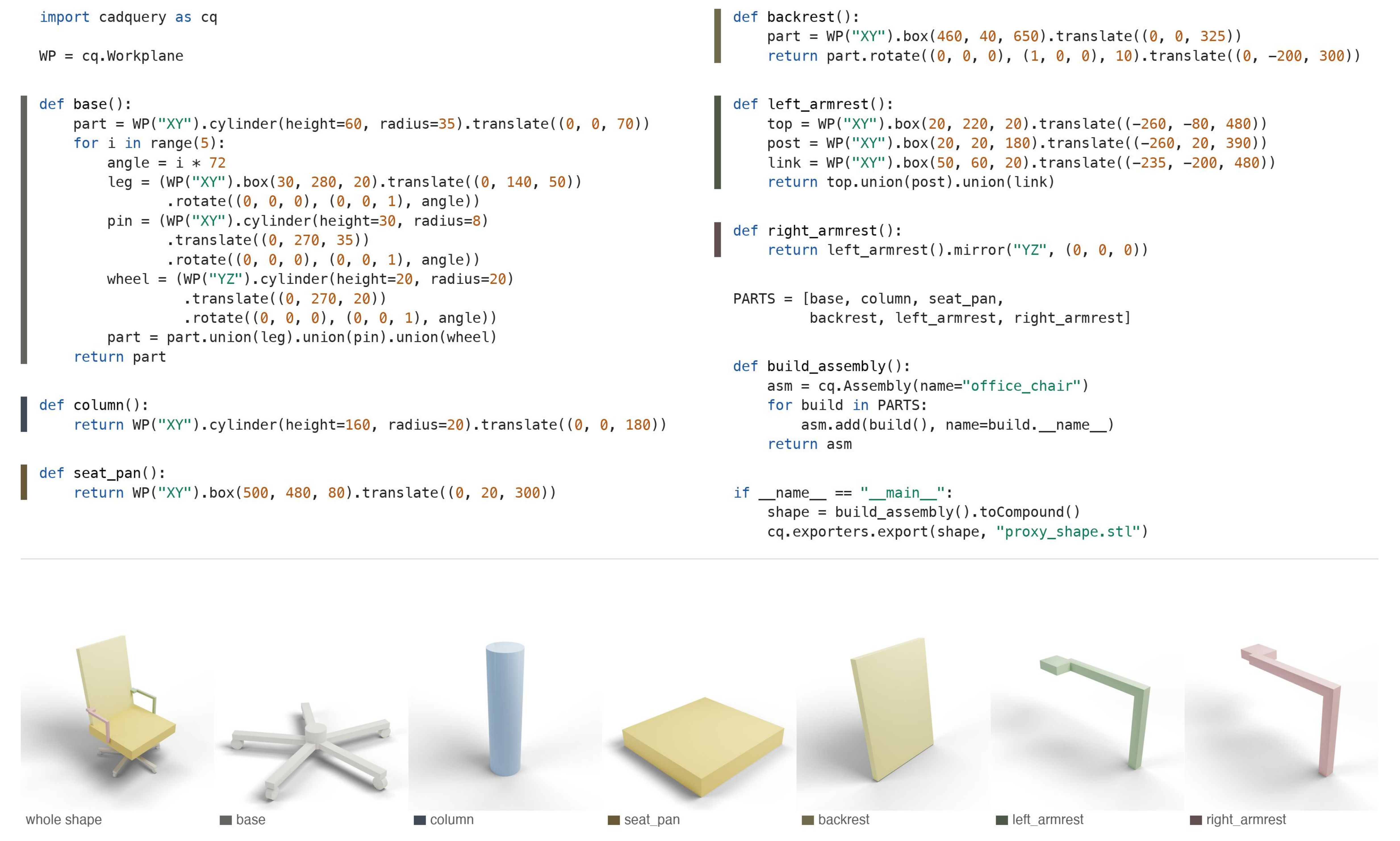}
\caption{The proxy program for the office chair. Each part is one function returning a simple volume, and the renders below show the parts and the assembled proxy.}
\label{fig:proxy_program}
\end{figure*}

\begin{figure*}[t!]
\includegraphics[width=\linewidth, height=20.5cm]{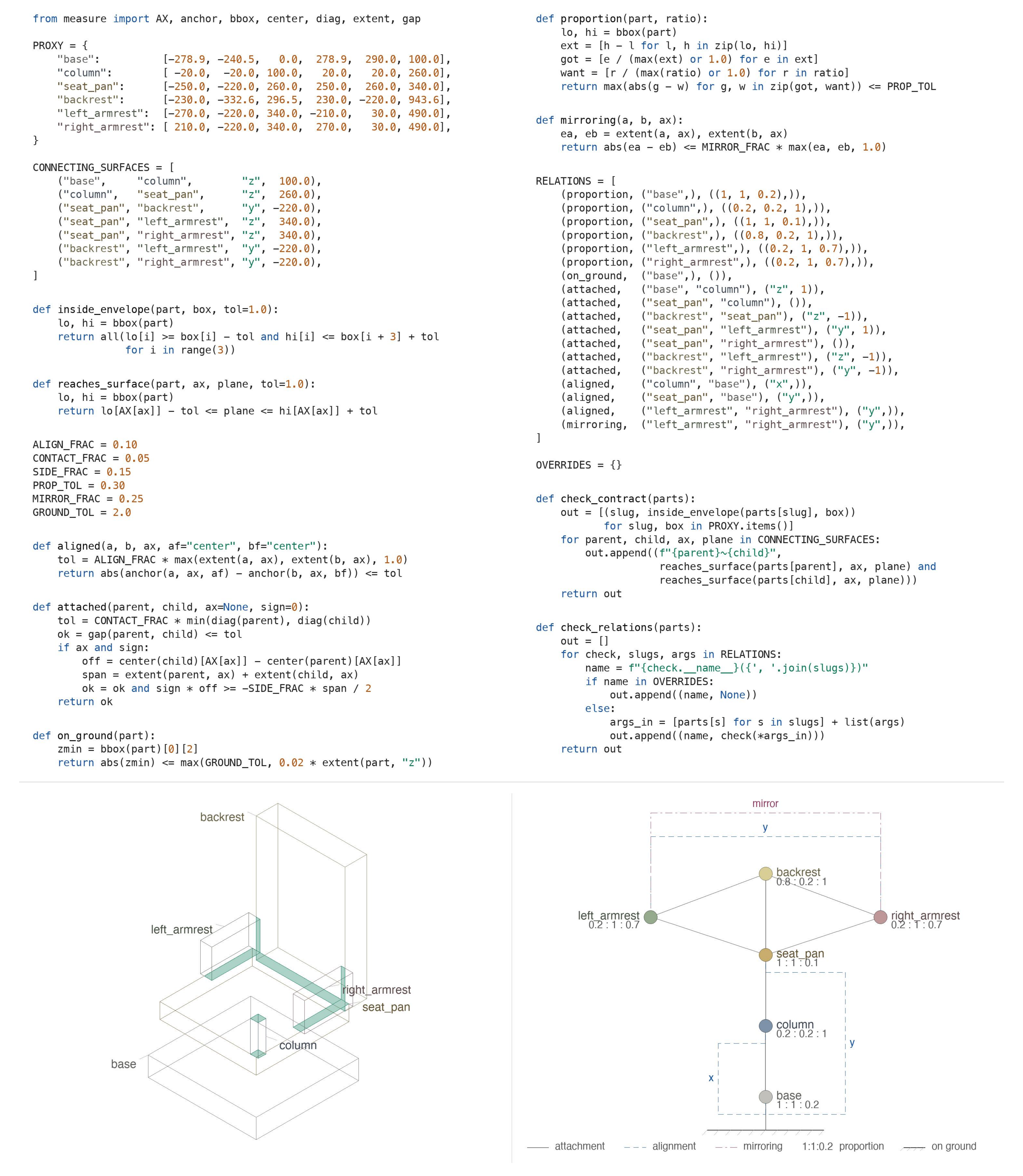}
\caption{The constraint program for the office chair. It holds the bounding boxes and connecting surfaces, the executable predicates that check each relation, and the extracted attachment, alignment, proportion, mirroring, and resting relations. Below: the connecting surfaces on the proxy (left) and the relation graph over the parts (right).}
\label{fig:constraint_program}
\end{figure*}

\begin{figure*}[t!]
\includegraphics[width=\linewidth, height=21cm]{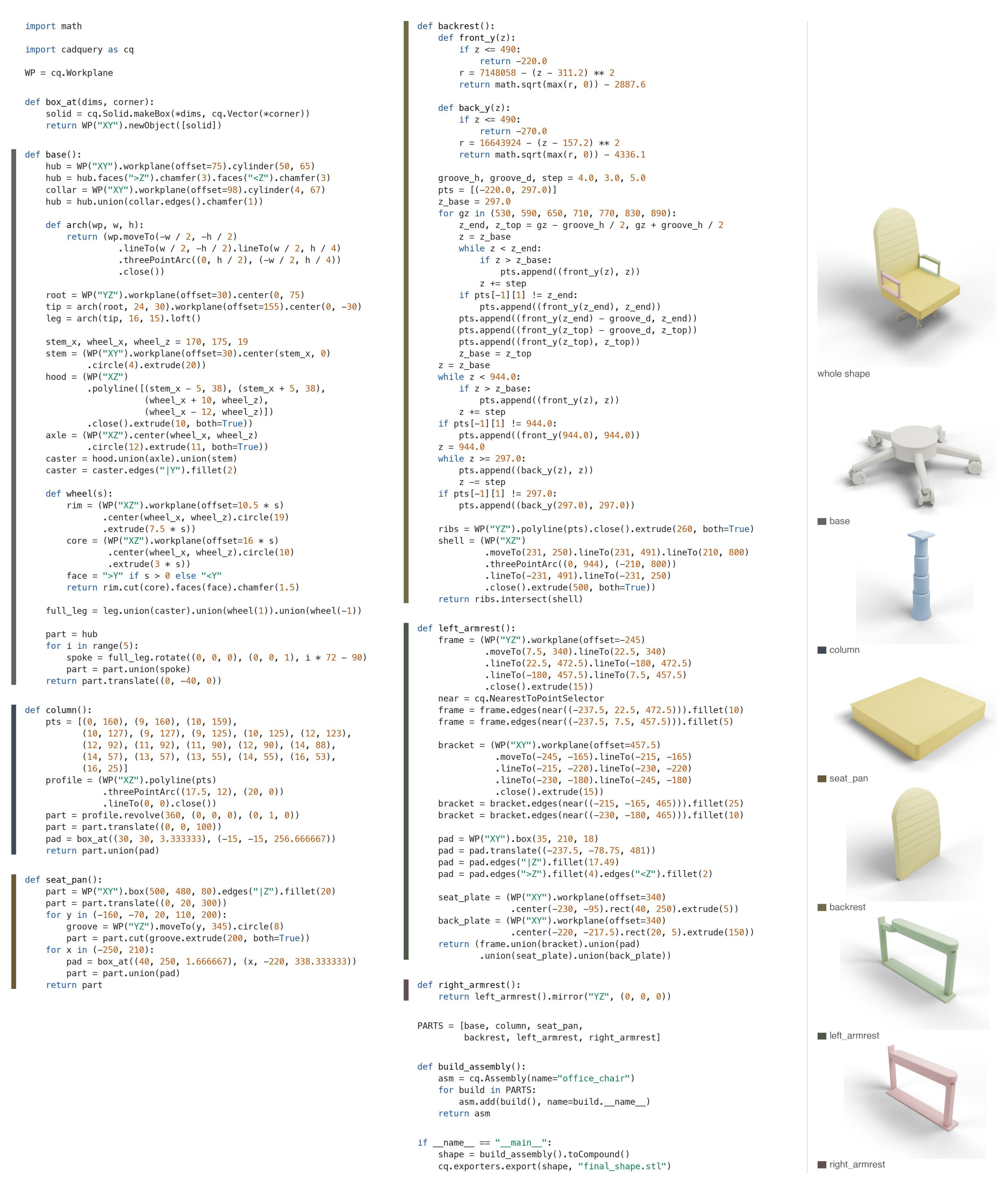}
\caption{The final program for the office chair, after detail generation and the deterministic rewrite. Each part is elaborated from its proxy volume into detailed geometry, with the renders on the right showing the parts and the assembled shape.}
\label{fig:final_program}
\end{figure*}

To give a concrete sense of what our system produces, we show the three artifacts it generates for one object, an office chair. The planning stage emits the proxy program (Figure~\ref{fig:proxy_program}) and the constraint program (Figure~\ref{fig:constraint_program}); detail generation and the rewrite produce the final program (Figure~\ref{fig:final_program}).

